\documentclass[iop,apj,appendixfloats,twocolappendix]{emulateapj}
\usepackage{apjfonts}
\usepackage{graphicx}
\usepackage{amsmath}
\usepackage{amssymb}
\usepackage{ulem}
\usepackage{soul}
\usepackage{float}
\usepackage[colorlinks=true,linkcolor=blue,citecolor=blue]{hyperref}%

\newcommand{\equ}[1]{eq.~(\ref{eq:#1})}

\newcommand{\se}[1]{\S\ref{sec:#1}}
\newcommand{\fig}[1]{Fig.~\ref{fig:#1}}
\newcommand{\figs}[1]{Figs.~\ref{fig:#1}}
\newcommand{\figss}[1]{\ref{fig:#1}}
\newcommand{\Fig}[1]{Figure~\ref{fig:#1}}

\newcommand{\tab}[1]{Table~\ref{tab:#1}}
\newcommand{\be}{\begin{equation}}
\newcommand{\ee}{\end{equation}}
\newcommand{\bea}{\begin{eqnarray}}
\newcommand{\eea}{\end{eqnarray}}

\newcommand{\no}{\noindent}

\newcommand{\Msun}{M_\odot}

\newcommand{\ifm}[1]{\relax\ifmmode#1\else$\mathsurround=0pt #1$\fi}
\newcommand{\kms}{\ifmmode\,{\rm km}\,{\rm s}^{-1}\else km$\,$s$^{-1}$\fi}

\newcommand{\Mpc}{\,{\rm Mpc}}
\newcommand{\kpc}{\,{\rm kpc}}
\newcommand{\pc}{\,{\rm pc}}
\newcommand{\Gyr}{\,{\rm Gyr}}

\newcommand{\Myr}{\,{\rm Myr}}
\newcommand{\K}{\,{\rm K}}
\newcommand{\cmc}{\,{\rm cm^{-3}}}
\newcommand{\cms}{\,{\rm cm^{-2}}}

\newcommand{\ltsima}{$\; \buildrel < \over \sim \;$}
\newcommand{\lsim}{\lower.5ex\hbox{\ltsima}}
\newcommand{\gtsima}{$\; \buildrel > \over \sim \;$}
\newcommand{\gsim}{\lower.5ex\hbox{\gtsima}}

\def\la{\langle}

\def\cmc{\,{\rm cm}^{-3}}
\def\cms{\,{\rm cm}^{-2}}

\def\M*{M_{\rm *}}
\def\Mv{M_{\rm v}}
\def\Rv{R_{\rm v}}

\def\cs{c_{\rm s}}

\def\NHI{N_{\rm HI}}

\newcommand{\angstrom}{\mbox{\normalfont\AA}}

\usepackage{xcolor}

\lefthead{Mandelker et al.}
\righthead{Shattering of Cosmic Sheets}
\slugcomment{Submitted to ApJ}

\begin{document}
\vspace{1mm}

\title{Thermal Instabilities and Shattering in the High-Redshift WHIM: Convergence Criteria and Implications for Low-Metallicity Strong HI Absorbers}


\author{Nir Mandelker\altaffilmark{1,2,3,4,5}, 
Frank C. van den Bosch\altaffilmark{4},
Volker Springel\altaffilmark{6},
Freeke van de Voort\altaffilmark{7},
Joseph N. Burchett\altaffilmark{8,9},
Iryna S. Butsky\altaffilmark{10},
Daisuke Nagai\altaffilmark{4,11},
S. Peng Oh\altaffilmark{12}
\vspace{8pt}}

\altaffiltext{1}
{corresponding author: nir\_mandelker@ucsb.edu}
\altaffiltext{2}
{Kavli Institute for Theoretical Physics, University of California, Santa Barbara, CA 93106, USA}
\altaffiltext{3}
{Racah Institute of Physics, The Hebrew University of Jerusalem, Jerusalem 91904, Israel}
\altaffiltext{4}
{Department of Astronomy, Yale University, PO Box 208101, New Haven, CT, USA}
\altaffiltext{5}
{Heidelberger Institut f{\"u}r Theoretische Studien, Schloss-Wolfsbrunnenweg 35, 69118 Heidelberg, Germany}
\altaffiltext{6}
{Max Planck Institute for Astrophysics, Karl-Schwarzschild-Stra{\ss}e 1, D-85748 Garching, Germany}
\altaffiltext{7}
{School of Physics \& Astronomy, Cardiff University, Queen's Building, The Parade, Cardiff CF24 3AA, UK}
\altaffiltext{8}
{Department of Astronomy , New Mexico State University, PO Box 30001, MSC 4500, Las Cruces, NM 88001}
\altaffiltext{9}
{University of California, Santa Cruz; 1156 High St., Santa Cruz, CA 95064, USA}
\altaffiltext{10}
{Astronomy Department, University of Washington, Seattle, WA 98195, USA}
\altaffiltext{11}
{Department of Physics, Yale University, PO Box 208101, New Haven, CT, USA}
\altaffiltext{12}
{University of California - Santa Barbara, Department of Physics, CA 93106-9530, USA}

%

\begin{abstract}
Using a novel suite of cosmological simulations zooming in on a Mpc-scale intergalactic sheet or ``pancake'' at $z\sim 3-5$, we conduct an in-depth study of the thermal properties and HI content of the warm-hot intergalactic medium (WHIM) at those redshifts. 
The simulations span nearly three orders of magnitude in gas-cell mass, from $\sim 7.7\times 10^6\Msun$ to $\sim 1.5\times 10^4\Msun$, one of the highest resolution simulations of such a large patch of the inter-galactic medium (IGM) to date, enabling us to analyze the convergence of salient properties with increasing resolution. At $z\sim 5$, a strong accretion shock develops around the main pancake following a collision between two smaller sheets. Gas in the post-shock region proceeds to cool rapidly, triggering thermal instabilities and the formation of a multiphase medium. We find neither the mass, nor the morphology, nor the distribution of HI in the WHIM to be converged, even at our highest resolution. Interestingly, the lack of convergence is more severe for the less dense, more metal-poor, intra-pancake medium (IPM) in between the filaments and far from any star-forming galaxies. As the resolution increases, the IPM develops a shattered structure, consisting of $\sim \kpc$ scale clouds which contain most of the HI. 
From our lowest to highest resolution, the covering fraction of metal-poor ($Z<10^{-3}Z_{\odot}$ ) Lyman-limit systems ($\NHI>10^{17.2}\cms$) in the IPM at $z \sim 4$ increases from 3 to 15 percent, while that of Damped Lyman-$\alpha$ Absorbers ($\NHI>10^{20}\cms$) with similar metallicity increases threefold, from 0.2 to 0.6 percent, with no sign of convergence. We find that a necessary condition for the formation of a multiphase, shattered structure is resolving the cooling length, $l_{\rm cool}=\cs t_{\rm cool}$, at $T\sim 10^5\K$. If this scale is unresolved, gas ``piles up'' at these temperatures and cooling to lower temperatures becomes very inefficient. We conclude that state-of-the-art cosmological simulations are still unable to resolve the multi-phase structure of the low-density IGM, with potentially far-reaching implications. 
\end{abstract} 
 

\section{Introduction} 
\label{sec:intro}







\smallskip
Only a small fraction of the baryons and heavy elements in the Universe are found in galaxies, accounting for both their stellar component and the dense gas that comprises the interstellar medium \citep[ISM, e.g.][]{Peeples14,Tumlinson17,Wechsler18}. Rather, the majority of baryons and metals reside in the circumgalactic medium (CGM), gas outside galaxies but within dark matter halos, and the intergalactic medium (IGM), gas outside dark matter halos. The IGM, CGM, and ISM are all intimately linked to galaxy evolution through cycles of gas accretion, star-formation, galactic outflows, and eventual re-accretion, collectively referred to as the cosmic baryon cycle \citep[e.g.][]{Putman12,McQuinn16,Tumlinson17}. The physical properties and chemical composition of the IGM and CGM thus offer valuable insight into processes related to galaxy formation and evolution. Moreover, the distribution of neutral hydrogen in the IGM, particularly at high-$z$, can be used to constrain cosmic reionization as well as structure formation and the nature of dark matter through studies of the Lyman-$\alpha$ (hereafter Ly$\alpha$) forest \citep[e.g.][]{Rauch98,Viel13,Lidz14,McQuinn16,Eilers18}). 

\smallskip
In recent decades, the diffuse gas in the CGM and the IGM have been probed using absorption line spectroscopy along lines of sight to distant QSOs or galaxies (e.g. \citealp{Burbidge68,Lynds71,Bergeron86,Hennawi06,Steidel10}; see also \citealp{Burchett20} for a more unconventional approach). Intervening gas clouds with neutral hydrogen column densities $\NHI\lsim 10^{15}\cms$ are understood to reside in the IGM and comprise the Ly$\alpha$ forest (see \citealp{Rauch98} and \citealp{McQuinn16} for reviews). 
Higher column density clouds, $\NHI> 10^{17.2}\cms$, are optically thick at wavelengths below the Lyman limit, $\lambda<$ 912\angstrom, and are thus referred to as Lyman Limit Systems (LLSs). These are commonly thought to reside in the CGM due to their large column densities, though several recent studies have postulated a growing population of low-metallicity LLSs with $Z<10^{-3}Z_{\odot}$ in the IGM at redshifts $z>2$ \citep{Fumagalli13,Robert19,M19}. New observational surveys such as the KODIAQ-Z survey \citep{OMeara15,OMeara21} are probing the metallicity distribution of strong HI absorbers, with $\NHI \sim (10^{15}-10^{19})\cms$ at redshifts $z>2$, in an effort to better understand the origin of dense, neutral gas and the transport and mixing of metals in the CGM and the IGM. Likewise, the advent of new integral field unit (IFU) spectographs such as KCWI on Keck and MUSE on the VLT have enabled emission line studies of the CGM and IGM around galaxies at similar redshifts \citep[e.g.][]{Steidel00,Cantalupo14,Martin14a,Martin14b,Leclercq17,Umehata19}. These observations reveal 
that the gas in and around galaxy halos has a complex multiphase structure, with cool clouds embedded in hotter ambient gas (see \citealp{Tumlinson17} for a recent review in the context of the CGM).

\smallskip
Using cosmological simulations to study the phase structure of the CGM and the IGM is notoriously difficult. Most state-of-the-art simulations employ a quasi-Lagrangian adaptive resolution, where the mass of resolution elements is kept fixed. The spatial resolution thus becomes very poor in the low density CGM and even worse in the IGM \citep{Nelson16}, orders of magnitude larger than the characteristic size of cool gas clouds in these systems. This is predicted to be the cooling length of $T\sim 10^4\K$ gas, $l_{\rm cool,min}=\cs t_{\rm cool}\sim 100\pc~(n/10^{-3}\cmc)^{-1}$, where $\cs$ is the sound speed and $t_{\rm cool}$ is the cooling time in the cool phase, assumed here to have $T\sim 10^4\K$ \citep{McCourt18,Sparre19,Das21}. To overcome these issues, several groups have recently introduced different methods to better resolve the CGM \citep{vdv19,Peeples19,Hummels19,Suresh19}. These studies have found that, despite no apparent systematic change to galaxy properties, the abundance, morphology, and distribution of cold, dense, low-ionization gas in the CGM is not converged, even with a fixed resolution of $500\pc$ throughout the CGM. For example, as the resolution is increased, the radial extent, covering fractions, and column densities of neutral hydrogen (HI) increase as well. \citet{Hummels19} suggest that this is due in part to increased numerical diffusion in low-resolution Eulerian simulations which leads to overmixing of cold and hot gas, and in part to higher resolution simulations better sampling high densities in a turbulent medium which leads to more efficient cooling. We discuss this in more detail in \se{disc}. However, different simulations disagree on the magnitude of the effect of enhanced CGM refinement, at least in part due to the different subgrid models employed by different groups for galaxy formation physics, such as stellar and AGN feedback, galactic winds, and gas photoheating and photoionization. This has obscured the details of why higher resolution leads to more cold gas, and what a meaningful convergence criterion might be.

\smallskip
Despite the success of recent enhanced refinement techniques for studying the CGM in simulations, the IGM remains very poorly resolved. Given its vast volume, the IGM does not readily lend itself to similar fixed-volume-refinement techniques that are being used for CGM studies. The enhanced refinement region in these simulations, as well as more standard ``zoom-in'' simulations, typically only extends to $\sim (1-2)\Rv$, with $\Rv$ the virial radius of the dark matter halo, leaving the vast majority of the IGM poorly resolved. This is in part due to a ``common wisdom'' that the diffuse IGM is a relatively simple system without very stringent convergence requirements. Previous studies have found that Ly$\alpha$ forest statistics, which are sensitive to gas clouds with HI column densities up to $\NHI\lsim 10^{15}\cms$, are converged at roughly percent levels in particle-based SPH simulations with gas particle masses of $m_{\rm gas} \lsim 2\times 10^5 \Msun$ \citep{Bolton09}, and in grid-based AMR simulations with cell sizes of $\Delta \lsim 20\kpc$ \citep{Lukic15}. Both of these studies found that the convergence requirements were more stringent at $z\sim (4-5)$ than at $z\lsim 2$. They postulated that this was because at low to moderate redshifts Ly$\alpha$ forest absorbers probe moderately overdense regions, while at high redshifts they probe underdense regions, since the neutral gas density in the IGM was overall higher.

\smallskip
By focusing on Ly$\alpha$ forest statistics, \citet{Bolton09} and \citet{Lukic15} limited their analysis to low density HI absorbers, with $\NHI\lsim 10^{15}\cms$. Both of these studies explicitly ignored higher column density absorbers such as LLSs, which they could not reliably model since 
their simulations did not include self-shielding of dense gas to the UV background. Furthermore, their simulations did not focus on the multiphase structure of the IGM or the so-called warm-hot intergalactic medium (WHIM). The latter is thought to dominate the filaments and sheets that comprise the cosmic-web of matter on large scales. Combined, these comprise between $\sim (25-50)\%$ of the volume and $\sim (50-75)\%$ of the mass in the Universe, depending on the method one uses to identify the various cosmic-web components, with sheets dominating the volume and filaments dominating the mass \citep[e.g.][and references therein]{Wang12,Cautun14,Libeskind18}. Modern cosmological simulations reveal strong accretion shocks around both intergalactic filaments \citep{Ramsoy21} and sheets \citep{M19}, similar to virial accretion shocks around massive dark-matter halos \citep[e.g.][]{Rees77,White78,bd03,Fielding17,Stern20}. At high redshift, the accretion shocks around filaments and sheets are predicted to be thermally unstable, leading to a cool core of $\sim 10^4\K$ gas surrounded by shocked 
gas with $T\sim (10^5-10^6)\K$ \citep{db06,Birnboim16}, 
which is indeed seen in cosmological simulations \citep{M19,Ramsoy21}. 

\smallskip
If intergalactic filaments and sheets are surrounded by thermally unstable accretion shocks, then it stands to reason that their phase structure, and the presence of dense gas in particular, is far from converged in simulations, similar to the CGM. 
Indeed, we have previously shown that with sufficient resolution, dense, metal-free LLSs can form in cosmic sheets far from any galaxy as a result of thermal instabilities
\citep{M19}, which may explain several recently observed absorption systems \citep{Fumagalli16,Lehner16,Robert19}. With current and upcoming surveys such as KODIAQ-Z aimed at studying the metallicity and spatial distribution of strong HI absorbers, understanding the formation and prevalence of such systems in the IGM is increasingly important. Furthermore, since these systems by their very nature are far from any galaxies, they offer us the opportunity to study thermal instabilities, convergence of gas thermal properties, and the formation of a multiphase medium in a cosmological context, without the complicating factors of galaxy-formation physics that affect CGM studies. This will allow us to gain insight into physical mechanisms which are also at play in the CGM, and better understand their convergence criteria.

\smallskip
In this paper, we use a suite of cosmological simulations first introduced in \citet{M19} to study the HI content, metal content, and gas thermal-phase structure in intergalactic sheets and filaments as a function of resolution, focusing on redshifts $z\sim (3-5)$. Our simulations zoom-in on a large region of the IGM between two massive galaxies at $z\sim 2.3$, with a comoving separation of $\sim 3\Mpc/h$. Our suite of four simulations span nearly 3 orders of magnitude in mass resolution, the best of which is one of the highest resolution simulations of such a large patch of the IGM to date. We describe our simulations in \se{sim}. In \se{results_1}, we describe the evolution of the large-scale structure and cosmic web in our simulations, and visually assess convergence of the morphology and distribution of HI and metals. In \se{results_2}, we perform a more quantitative convergence study of the distribution and morphology of HI, in particular the presence of strong HI absorbers and large gas clumping factors. In \se{results_3} we focus on the thermal properties and phase-structure of very metal-poor gas in sheets, far from any galaxies. In \se{disc} we discuss the physical mechanisms behind the convergence (or lack thereof) of the gas thermal properties, and compare our results to recent studies of the multiphase CGM. Finally, we conclude in \se{conc}. Throughout, we assume a flat $\Lambda$CDM cosmology with $\Omega_{\rm m}=1-\Omega_{\Lambda}=0.3089$, $\Omega_{\rm b}=0.0486$, $h=0.6774$, $\sigma_8=0.8159$, and $n_{\rm s}=0.9667$ \citep{Planck16}.

\section{Simulation Method} 
\label{sec:sim}

\smallskip
We use the quasi-Lagrangian moving-mesh code \texttt{AREPO} \citep{Springel10} for our simulations. To select our target region, we first consider the 200 most massive halos at $z\sim 2.3$ in the Illustris TNG100\footnote{http://www.tng-project.org} magnetohydrodynamic cosmological simulation \citep{Pillepich18b,Nelson18,Springel18}. These span a mass range of $\Mv\sim (1.0-40)\times 10^{12}\Msun$, where $\Mv$ is the virial mass defined using the \citet{Bryan98} spherical overdensity. We then select all halo pairs with a comoving separation in the range $(2.5-4.0)\Mpc/h \sim (3.7-5.9)\Mpc$. We find 48 such halo pairs, each connected by the cosmic web, either lying in the same cosmic sheet or connected by a cosmic filament. We selected one pair at random, consisting of two halos with $\Mv\sim 5\times 10^{12}\Msun$ each, separated by a proper distance of $D\sim 1.2\Mpc$. By $z=0$, the two halos evolve into mid-size groups with $\Mv\sim (1.6-1.9)\times 10^{13}\Msun$, separated by $\sim 2.7\Mpc$. Their comoving distance has thus decreased by $\lsim 30\%$. 

\smallskip
We define $R_{\rm ref}=1.5\times R_{\rm v,max}\sim 240\kpc$, with $R_{\rm v,max}$ the larger of the two virial radii at $z=2.3$. The zoom-in region is the union of a cylinder with radius $R_{\rm ref}$ and length $D$ extending between the two halo centers, and two spheres of radius $R_{\rm ref}$ centred on either halo. We trace all dark matter particles within this volume back to the initial conditions of the simulation, at $z=127$, refine the corresponding Lagrangian region to higher resolution, and rerun the simulation to a final redshift, $z_{\rm fin} \gsim 2$ (see \tab{sims}), when the region of interest by construction becomes contaminated by low resolution material from outside the refinement region. The simulations were performed with the same physics model as used in the TNG100 simulation, described in detail in \citet{Weinberger17} and \citet{Pillepich18}. We briefly summarize below the implementation of the ionizing radiation field and of cooling, and our method for identifying dark matter halos, which are most relevant to our current work. 

\begin{table}
\centering
\begin{tabular}{c|c|c|c|c|c}
\hline
Sim. Name & $m_{\rm dm}~[\Msun]$ & $\epsilon_{\rm dm}~[\pc]$ & $m_{\rm gas}~[\Msun]$ & $\epsilon_{\rm gas}~[\pc]$ & $z_{\rm fin}$ \\
\hline
ZF0.5 & $4.2\times 10^7$ & 2000 & $7.7\times 10^6$ & 1000 & 1.94 \\
ZF1.0 & $5.2\times 10^6$ & 1000 & $9.6\times 10^5$ & 500  & 1.94 \\
ZF2.0 & $6.5\times 10^5$ & 500  & $1.2\times 10^5$ & 250  & 1.94 \\
ZF4.0 & $8.2\times 10^4$ & 250  & $1.5\times 10^4$ & 125  & 2.91 \\
\hline
\hline
TNG300 & $5.9\times 10^7$ & 1480 & $1.1\times 10^7$ & 370 & 0.0 \\
TNG100 & $7.5\times 10^6$ &  740 & $1.4\times 10^6$ & 185 & 0.0 \\
TNG50  & $4.5\times 10^5$ &  288 & $8.5\times 10^4$ &  74 & 0.0 \\
\hline
\end{tabular}
\caption{\textit{Top four rows:} Parameters of the simulations studied in this work. From left to right, the columns list the simulation name given as its Zoom Factor, ZF (see text), the dark matter particle mass, $m_{\rm dm}$ in $\Msun$, the Plummer-equivalent gravitational softening of the dark matter particles, $\epsilon_{\rm dm}$ in $\pc$, the target gas cell mass, $m_{\rm gas}$ in $\Msun$, the minimal gravitational softening for gas cells, $\epsilon_{\rm gas}$ in $\pc$, and the final redshift of the simulation, $z_{\rm fin}$. In the \textit{bottom three rows,} we list for comparison the corresponding values from the three flagship simulations of the Illustris TNG suite, taken from table 1 of \citet{Nelson19}.}
\vspace{-13pt}
\label{tab:sims}
\end{table}

\smallskip
We follow the production and evolution of nine elements (H, He, C, N, O, Ne, Mg, Si, and Fe). These are produced in supernovae Type Ia and Type II and in AGB stars according to tabulated mass and metal yields. The neutral hydrogen fraction is calculated on the fly, using the ionization and recombination rates from \citet{Katz96} and the ionizing ultraviolet background (UVB) from \citet{FG09}, which is instantaneously switched on at $z=6$ and is assumed to be spatially uniform but redshift dependent. To minimize any potential influence of this instantaneous switching on of the UVB, we limit our current analysis to $z\le 5$. Self-shielding from the UV background is implemented using fits to the degree of self-shielding as a function of hydrogen volume density and redshift in radiative transfer simulations following \citet{Rahmati13}. Metal line cooling is included using pre-calculated rates as a function of density, temperature, metalicity and redshift \citep{Wiersma09}, with corrections for self-shielding. Cooling is further modulated by the radiation field of nearby active galactic nuclei (AGN) by superimposing the UVB with the AGN radiation field within $3\Rv$ of halos containing actively accreting supermassive black holes \citep{Vogelsberger13}. Gas with densities greater than $n_{\rm thresh}=0.13\cmc$ is considered eligible for star-formation and is placed on an artificial equation of state meant to mimic the unresolved multiphase ISM \citep{Springel03}. Its temperature thus does not represent the gas thermal temperature. However, our analysis will focus almost exclusively on gas with densities below the star formation threshols, $n<n_{\rm thresh}$.

\smallskip
To identify dark matter halos in the simulation, we apply the same procedure as implemented in the Illustris TNG simulations. Namely, we first apply a Friends-of-Friends (FoF) algorithm with a linking length $b = 0.2$ to the dark matter particles, then assign gas and stars to FoF groups based on their nearest-neighbour dark matter particle, and finally apply \texttt{SUBFIND} \citep{Springel01,Dolag09} to the total mass distribution in each FoF group. The most massive \texttt{SUBFIND} object in each FoF group is considered the central halo, its virial radius $\Rv$ is defined using the \citet{Bryan98} spherical overdensity criterion, and its virial mass $\Mv$ is the total mass of dark matter, gas, and stars within $\Rv$. 

\begin{figure*}
\begin{center}
\includegraphics[trim={0.1cm 0.2cm 0.0cm 0.2cm}, clip, width =0.98 \textwidth]{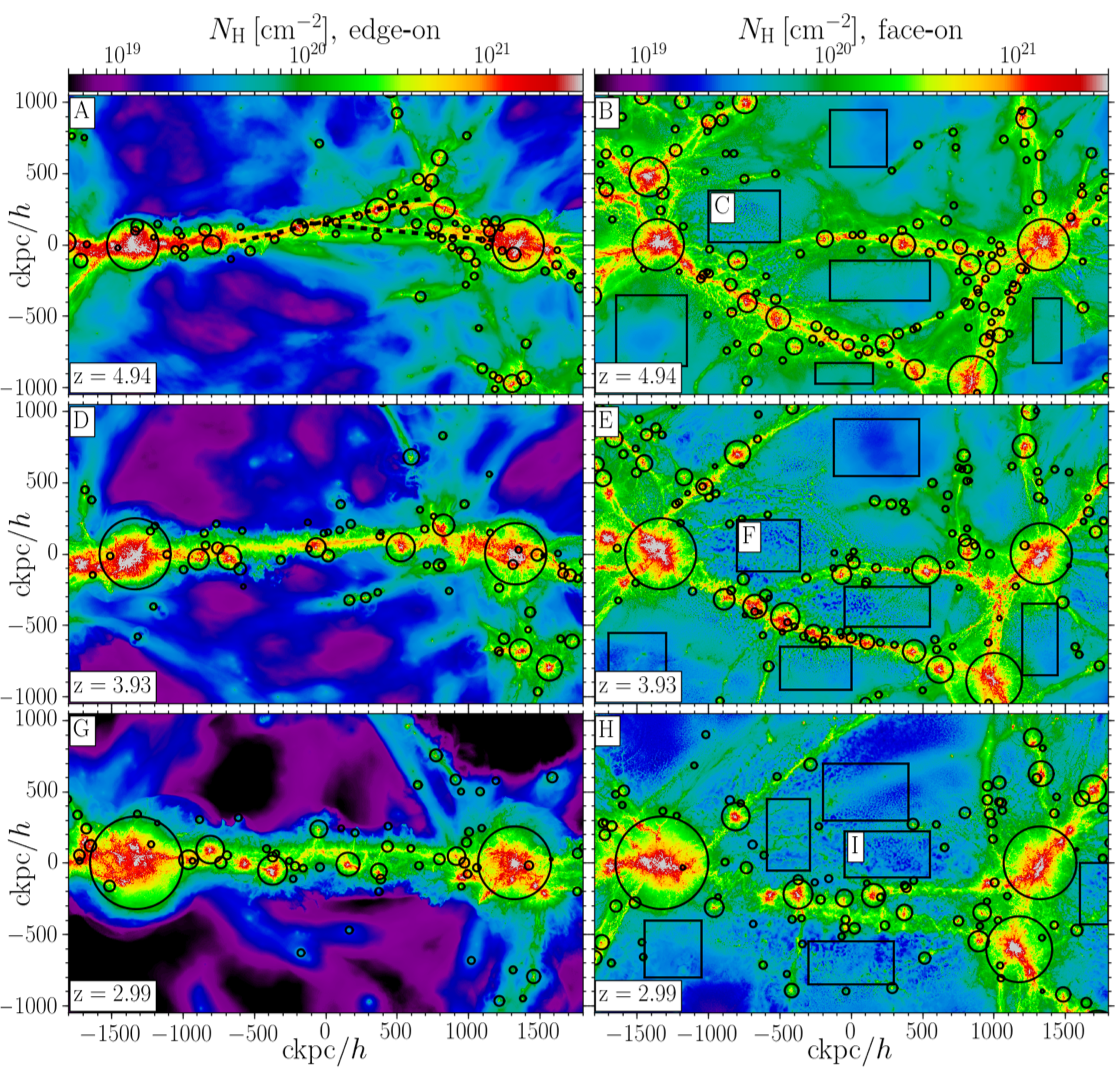}
\end{center}
\caption{The zoom-in region of the ZF4.0 simulation. We show the large scale structure surrounding the two main halos at $z\sim 5$ (top), $z\sim 4$ (middle) and $z\sim 3$ (bottom). The two halos lie in a cosmic sheet, shown edge-on and face-on in the left- and right-hand columns, respectively. Black rectangles in the right-hand column mark individual sheet regions that are explored in more detail and referenced throughout the text. The color scale indicates the total hydrogen column density, $N_{\rm H}$, integrated over $\pm 400~{\rm ckpc}/h$. Black circles mark central dark matter halos with virial mass $\Mv>10^9\Msun$, while their sizes denote $\Rv$. Nearly all these halos lie along dense filaments within the sheet. The main sheet is formed by an oblique collision between two smaller sheets at $z\sim 5$, marked with dashed lines in panel A. Following the collision, the column-density of sheet gas displays small-scale structure and fluctuations.}
\label{fig:NH_large}
\end{figure*}

\begin{figure*}
\begin{center}
\includegraphics[trim={1.0cm 1.0cm 2.9cm 0.8cm}, clip, width =0.98 \textwidth]{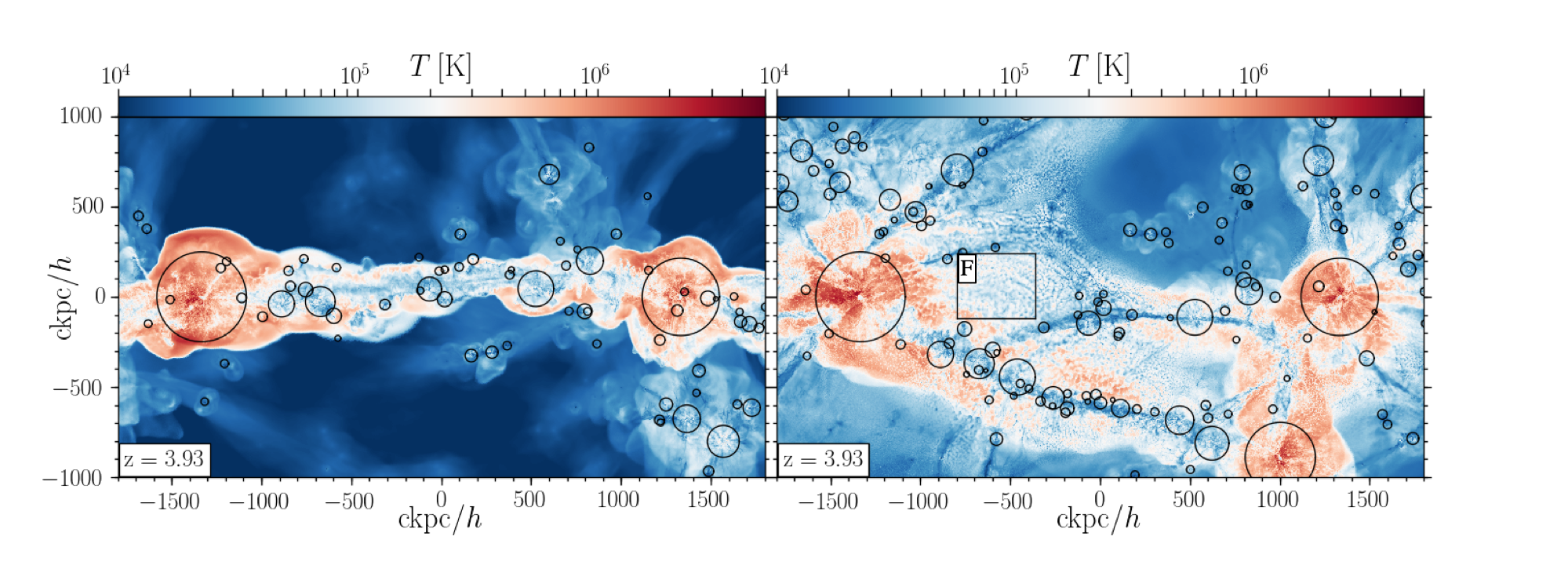}
\end{center}
\caption{Temperature map of the ZF4.0 simulation at $z=4$, in the same frame as panels D and E in \fig{NH_large}. Black circles show central halos with $\Mv>10^9\Msun$ as in \fig{NH_large}, while the black rectangle highlights region F from panel E in \fig{NH_large}. We show the density-weighted average temperature along the line of sight, integrated over $\pm 400~{\rm ckpc}/h$. The collision between the two inclined sheets at $z\sim 5$ generates a strong shock at the sheet edge, visible in the edge-on view. In the face-on view, the filaments appear cold while the sheet regions have a multiphase structure with hot and cold regions owing to thermal instability in the post-shock medium.}
\label{fig:Temp_large}
\end{figure*}

\smallskip
We performed five simulations with different resolution within the refinement region. In this work we focus on four of these simulations, listed in \tab{sims}, each characterised by a ``Zoom-Factor'', or ZF. Our fiducial simulation, ZF1.0, has a dark matter particle mass of $m_{\rm dm}=5.2\times 10^6\,\Msun$ and a Plummer-equivalent gravitational softening of $\epsilon_{\rm dm}=1000\pc$ comoving. Gas cells are refined such that their mass is within a factor of 2 of the target mass, $m_{\rm gas}=9.6\times 10^5\Msun$. Gravitational softening for gas cells is twice the cell size, down to a minimal gravitational softening $\epsilon_{\rm gas}=0.5\epsilon_{\rm dm}=500\pc$. Thus, ZF1.0 has slightly better mass resolution and slightly worse force-resolution than Illustris TNG-100. Additional simulations are labelled ZF0.5, ZF2.0, and ZF4.0, where the mass (spatial) resolution in ZF$x$ is $x^3$ ($x$) times better than ZF1.0. Thus, the resolution in ZF0.5 and ZF2.0 is comparable to Illustris TNG300 and TNG50 respectively (\tab{sims}). Our highest resolution simulation, ZF4.0, has $\sim 8$ times better mass resolution than TNG50\footnote{While the minimal gravitational softening for gas cells is slightly larger in ZF4.0 than in TNG50, this has no effect on our results for two reasons. First, in the IGM regions we focus on here the minimal cell size in ZF4.0 is $\gsim 100\pc$, so the gravitational softening is twice the cell size and larger than the minimal value anyway. Second, as we will show, the small scale clouds that form in high-resolution simulations are $\sim \kpc$-scale, not self-gravitating, and far enough away from galaxies to be minimally affected by resolution effects in the ISM. It is thus the mass resolution, and by association the cell size, which is the determinant resolution factor in our analysis.}, with $m_{\rm gas}=1.5\times 10^4\Msun$ and $m_{\rm dm}=8.2\times 10^4\Msun$. 
An additional simulation, ZF3.0, was performed but is not included in our current analysis, as it is simply midway between ZF2.0 and ZF4.0.

\section{Large Scale Structure Evolution and Morphology: Visual Inspection} 
\label{sec:results_1}

\smallskip
We begin in this section by examining the evolution of the large scale structure and the cosmic web in our simulations, from $z\sim (5-3)$ \citep[see also][]{M19}. 
We then examine the distribution of hydrogen, HI, and metals and preliminarily assess their convergence through visual inspection before moving on to more quantitative analysis in \se{results_2}. In what follows we focus on $z \sim 4$, but we note that results at $z \sim 3$ and $z\sim 5$ are qualitatively very similar.

\smallskip
In \fig{NH_large} we show the total hydrogen column density, $N_{\rm H}$, in ZF4.0 at $z\sim 5$ (top), $4$ (middle), and $3$ (bottom), in two orthogonal projections, with the intergalactic sheet containing the two halos shown edge-on (left) and face-on (right). At $z>5$ the system consists of two lower-mass sheets initially inclined to one another, marked by dashed lines in panel A. These merge at $z\sim 5$, leaving only a single sheet visible in panels D and G. The sheet contains several prominent co-planar filaments, with end-points at either of the two main halos and along which nearly all halos with $\Mv>10^9\Msun$ are located. Most of these filaments merge at $z<3$, leaving behind the single giant filament selected at $z=2.3$. The beginning of this merger is visible in panel H. 

\smallskip
In \fig{Temp_large} we show the projected, density-weighted gas temperature in the same frame as panels D and E from \fig{NH_large}. A planar accretion shock around the sheet, triggered by the earlier collision at $z\sim 5$, is clearly visible in the edge-on view, as are spherical accretion shocks around the two main halos. In the face-on view, the filaments appear cold, with $T\sim 2\times 10^4\K$, while the regions between the filaments exhibit a multiphase structure, with hot and cold gas coexisting in a granular structure. This same granular structure is visible in the gas column density (\fig{NH_large}). As the merging sheets were initially inclined, the collision and resulting granular structure propagate from left-to-right, as can be seen by comparing panels B and E in \fig{NH_large}. We hereafter refer to this gaseous medium in between the filaments and outside all dark matter halos with $\Mv>10^9\Msun$ as the \textit{Intra-Pancake Medium}, or IPM. In panels B, E, and H in \fig{NH_large} we highlight several IPM regions, labelled C, F, and I, which we analyze and discuss individually 
in \figs{nH_slice} and \figss{profiles}. 

\begin{figure*}
\begin{center}
\includegraphics[trim={0.5cm 0.0cm 0.0cm 0.0cm}, clip, width =0.96 \textwidth]{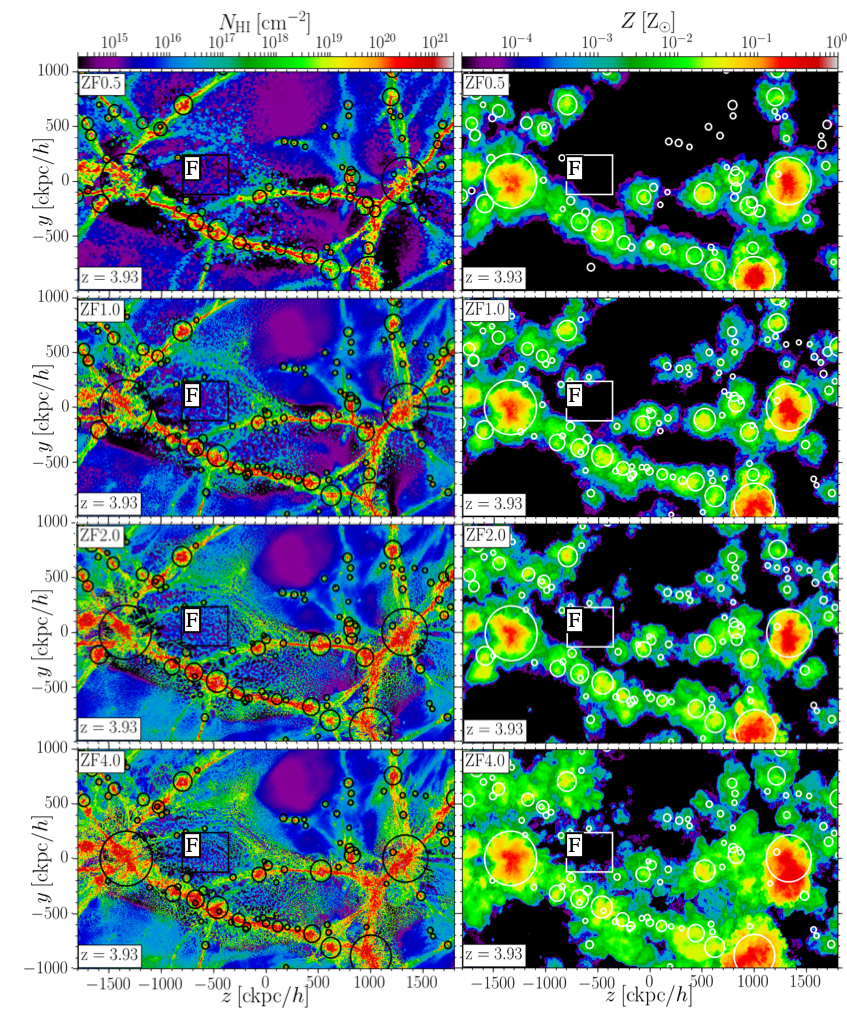}
\end{center}
\vspace{-10.0pt}
\caption{
Large scale distribution of HI and metals. Each panel represents the zoom-in region of the simulation at $z\sim 4$, integrated over $\pm 100~{\rm pkpc}\sim \pm 335~{\rm ckpc}/h$ from the sheet midplane. We show our four simulations, ZF0.5, ZF1.0, ZF2.0, and ZF4.0, with resolution increasing from top to bottom. Circles and rectangles are the same as in \fig{Temp_large} and in panel E of \fig{NH_large}. \textit{Left:} The HI column density, $\NHI$. While the morphology of the large-scale structure and the positions of dark matter halos are very similar at all resolutions, $\NHI$ increases markedly both in overall normalization and in the amplitude of fluctuations as the simulation resolution is increased. \textit{Right:} Mass-weighted average metallicity along the line-of-sight. In all simulations, the filaments are enriched to $[Z]\equiv {\rm log}(Z/Z_{\odot})\lsim -2.0$, while the IPM has much lower metalicity values as it is further away from star-forming galaxies. However, as the resolution increases, metals are distributed further and further into the IPM and away from filament spines. Nonetheless, in all cases filaments and the IPM can be roughly distinguished using a metallicity threshold of $[Z]_{\rm thresh}=-3.0$.}
\label{fig:NHI_large}
\end{figure*}

\begin{figure*}
\begin{center}
\includegraphics[trim={0.5cm 0.0cm 0.0cm 0.2cm}, clip, width =0.96 \textwidth]{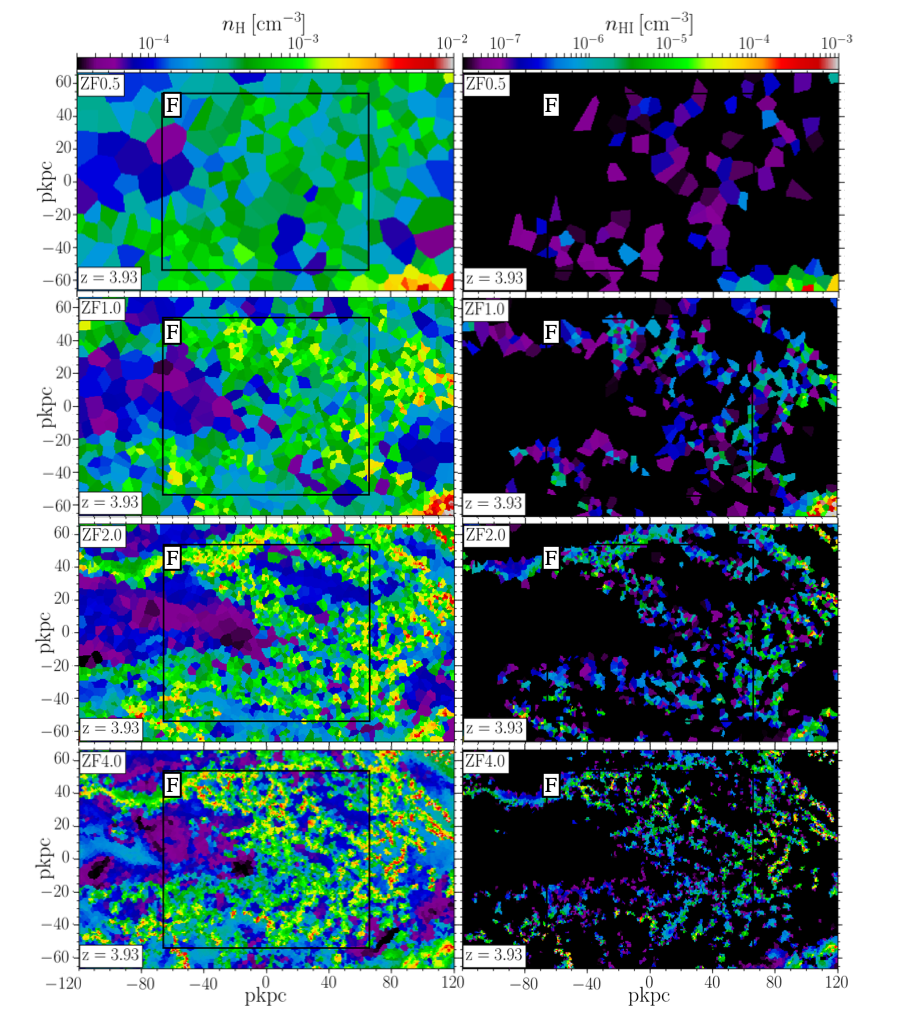}
\end{center}
\vspace{-10.0pt}
\caption{
Small scale structure in the IPM. Each panel represents an infinitesimally thin slice near the midplane of the sheet at $z=4$, in region F from \figs{NH_large}-\figss{NHI_large}. Resolution increases from top to bottom, as in \fig{NHI_large}. \textit{Left:} hydrogen number density, $n_{\rm H}$. In ZF0.5, the gas density is nearly constant at $n_{\rm H}\sim (2-5)\times 10^{-4}\cmc$ with only minor fluctuations. As the resolution increases, the gas seems to ``shatter'' into smaller and smaller clouds with higher and higher densities, embedded in a low density background. \textit{Right:} Neutral hydrogen number density, $n_{\rm HI}$. The HI clouds correspond to the dense small-scale clouds seen in the density distribution. As the resolution increases, these small-scale clouds grow denser and more prominent and lead to more HI in the IPM.
}
\label{fig:nH_slice}
\end{figure*}

\smallskip
In \fig{NHI_large} we explore the large-scale distribution of neutral hydrogen and metals in our zoom-in region as a function of resolution. In the left-hand column, we show the HI column density, $\NHI$, in a face on projection through the sheet at $z\sim 4$, integrated over $\pm 100{\rm pkpc}\sim 335{\rm ckpc}/h$. Note that this is slightly smaller than the $\pm 400{\rm ckpc}/h$ used in \figs{NH_large} and \figss{Temp_large}, but it is thick enough to contain all the HI gas in the vicinity of the sheet. We show projections from our four simulations, with resolution increasing from ZF0.5 (top) to ZF4.0 (bottom). The overall morphology of various cosmic-web elements, such as filaments and $\Mv>10^9 \Msun$ halos, are nearly identical at all resolutions. At all resolutions, filaments are dominated by very large values of $\NHI\gsim 10^{20}\cms$, classifying them as DLAs or sub-DLAs\footnote{A Damped Lyman-$\alpha$ Absorber, or DLA, has $\NHI>10^{20.3}\cms$ yielding an absorption line profile dominated by the damping wings of the Lorentzian part of the Voigt profile.}, while the IPM has $10^{14.5} \cms \lsim \NHI \lsim 10^{20} \cms$. However, both the average value and the level of fluctuations of $\NHI$ in the IPM noticeably increase with resolution. In ZF4.0, these fluctuations correspond to those seen in $N_{\rm H}$ (\fig{NH_large}) and temperature (\fig{Temp_large}). Column density values of $\NHI\sim (10^{18}-10^{19})\cms$ seem common throughout the IPM in ZF4.0, while in ZF0.5 and ZF1.0 these 
are largely restricted to the filaments.

\smallskip
In the right-hand column of \fig{NHI_large}, we show the mass-weighted average metallicity along the line of sight, averaged over $\pm 100{\rm pkpc}$  in the same face-on projection through the sheet. This corresponds to the total metal mass divided by the total gas mass in every column. In all cases, the filaments are enriched to $[Z]\equiv {\rm log}(Z/Z_{\odot}) \sim -2.0$ by $z\sim 4$, while the IPM has significantly lower metallicity values. This is unsurprising, since most star-forming galaxies and nearly all halos with $\Mv>10^9\Msun$ are located along the filaments. Galactic outflows thus enrich intergalactic filaments with metals at very high redshift, while the IPM, which is nearly devoid of star-forming galaxies, remains pristine. However, we see that in higher resolution simulations the metals are distributed over a larger volume and reach larger distances from the filament spines. Thus, in addition to being unconverged in terms of HI content, the WHIM and in particular the IPM are unconverged in terms of their metal content. There are three potential reasons for this. First, higher resolution simulations resolve star-formation in lower mass galaxies and at earlier times, thus allowing more metals to propagate into the IGM for a longer time. Second, at a given halo mass, higher resolution simulations resolve higher star formation rates, and thus in turn launch more powerful winds, delivering metals to larger distances from galaxies. Third, higher resolution simulations better resolve turbulence and turbulent metal mixing in the IGM. Quantifying the relative importance of these three mechanisms is beyond the scope of this paper, as we instead focus on the production and distribution of cold, dense gas and neutral hydrogen in the IGM. 

\smallskip
In \fig{nH_slice}, we explore the small-scale structure of the IPM. In the left-hand column, we show the hydrogen density in an infinitesimally thin slice near the midplane of the sheet at $z\sim 4$, in region F from \figs{NH_large}-\figss{NHI_large}. As in \fig{NHI_large}, resolution increases from ZF0.5 on the top to ZF4.0 on the bottom. The difference in gas morphology between the simulations is striking. In ZF0.5, the gas density exhibits only minor fluctuations around a typical value of $\sim (2-5)\times 10^{-4}\cmc$. As the resolution increases, more and more small-scale structure appears in the form of dense cloudlets. This is reminiscent of thermal ``shattering'' as described by \citet{McCourt18}. According to this picture, non-linear thermal instabilities in a cooling, pressure-confined medium cause the medium to fragment, or ``shatter'', into dense cloudlets with $T\gsim 10^4\K$ in pressure equilibrium with a more tenuous, hot background. The size of these cloudlets is set by the \textit{cooling length}, 
\be
\label{eq:lcool}
l_{\rm cool}=\cs t_{\rm cool}=\left[\frac{\gamma\,k_{\rm B}^3}{(\gamma-1)^2\,\mu\,m_{\rm p}}\right]^{1/2}\frac{T^{3/2}}{n\Lambda(T)},
\ee
{\no}where $\gamma$ is the adiabatic index of the gas and is $5/3$ for an ideal monoatomic gas, $k_{\rm B}$ is Boltzmann's constant, $m_{\rm p}$ is the proton mass, $\mu$ is the mean molecular weight of the gas and is $\sim 0.59$ for a fully ionized gas of primordial composition, T is the gas temperature, $n$ is the particle number density, and $\Lambda(T)$ is the temperature dependent cooling function. This is the largest lengthscale that can maintain pressure equilibrium over a cooling time. Shattering is hierarchical, in the sense that as the gas cools $l_{\rm cool}$ decreases, causing existing cloudlets to shatter into even smaller cloudlets, in much the same way as gravitational Jeans instability can lead to hierarchical fragmentation. This process continues until the \textit{minimum cooling length} is reached, typically near the hydrogen peak of the cooling curve at $T\gsim 10^4\K$. For isobaric cooling, $n\propto T^{-1}$, and $l_{\rm cool,min}\sim 100\pc~(n/10^{-3}\cmc)^{-1}$ for gas in collisional ionization equilibrium, where $n$ is the density at $T\gsim 10^4\K$ \citep{McCourt18}. For the UVB assumed in our simulations\footnote{Note that \citet{McCourt18} did not assume a UVB but rather a cooling floor at $10^4\K$. This causes $l_{\rm cool,min}$ in our simulations to be slightly larger than in their estimates, since the UVB alters the cooling curve.} $l_{\rm cool}$ is minimal at $T\sim 2\times 10^4\K$ at typical IPM densities. At densities of $n_{\rm H}\sim 10^{-2.5}\cmc$, typical of the dense cloudlets in ZF4.0, $l_{\rm cool,\,min}\sim 1\kpc$ is comparable to the cloudlets' size \citep{M19}. 
For comparison, the typical (minimal) cell size in 
region F is $\Delta\sim 0.8~(0.3)\kpc$ in ZF4.0, and $\Delta\sim 4.0~(2.5)\kpc$ in ZF1.0. We discuss the shattering picture in the context of our system in more detail in \se{disc}. In particular, we discuss whether resolving $l_{\rm cool,min}$ is necessary for the formation of dense cloudlets, since this scale is unresolved in ZF2.0 and is at best marginally resolved in ZF4.0, despite both of these simulations exhibiting a similar shattered structure. 

\smallskip
We note that the thermal Jeans length in the cold phase is $L_{\rm J}=[9c_{\rm s}^2/(4\pi G\rho)]^{1/2}\sim 30\kpc$, significantly larger than the cooling length, the cloud sizes, and the typical cell size. This implies that the clouds are not the result of gravitational instability in the sheet, and that the lack of cloudlets in low resolution simulations is not a result of increased gravitational softening or decreased force resolution. Rather, this supports our hypothesis that they result from thermal instabilities.

\smallskip
In the right-hand column of \fig{nH_slice}, we show the neutral hydrogen density, $n_{\rm HI}$, in the same slice. Unsurprisingly, the HI is located in the dense cloudlets seen in the left column. This explains the enhancement of $\NHI$ with resolution seen in \fig{NHI_large} - higher resolution simulations better resolve the formation of small-scale dense clouds, possibly via ``shattering'' (see \se{disc}), thus enabling the formation of more neutral gas. We note that there is an additional runaway effect due to the implementation of self-shielding in the simulations following \citet{Rahmati13}. At densities $n_{\rm H}\sim 10^{-3}$ and $10^{-2}\cmc$, the UVB is roughly $\sim 90\%$ and $10\%$ of its unshielded value, respectively. Thus, as gas cools and its density increases beyond $10^{-3}\cmc$, 
the typical density 
in ZF0.5, the UVB is rapidly shielded and HI formation is enhanced. We address the impact of self-shielding on our results in \se{disc}.

\section{HI Mass Fractions and Covering Fractions, and Gas Clumping Factors} 
\label{sec:results_2}

\begin{figure*}
\begin{center}
\includegraphics[trim={0.4cm 0.0cm 4.0cm 0.0cm}, clip, width =0.98 \textwidth]{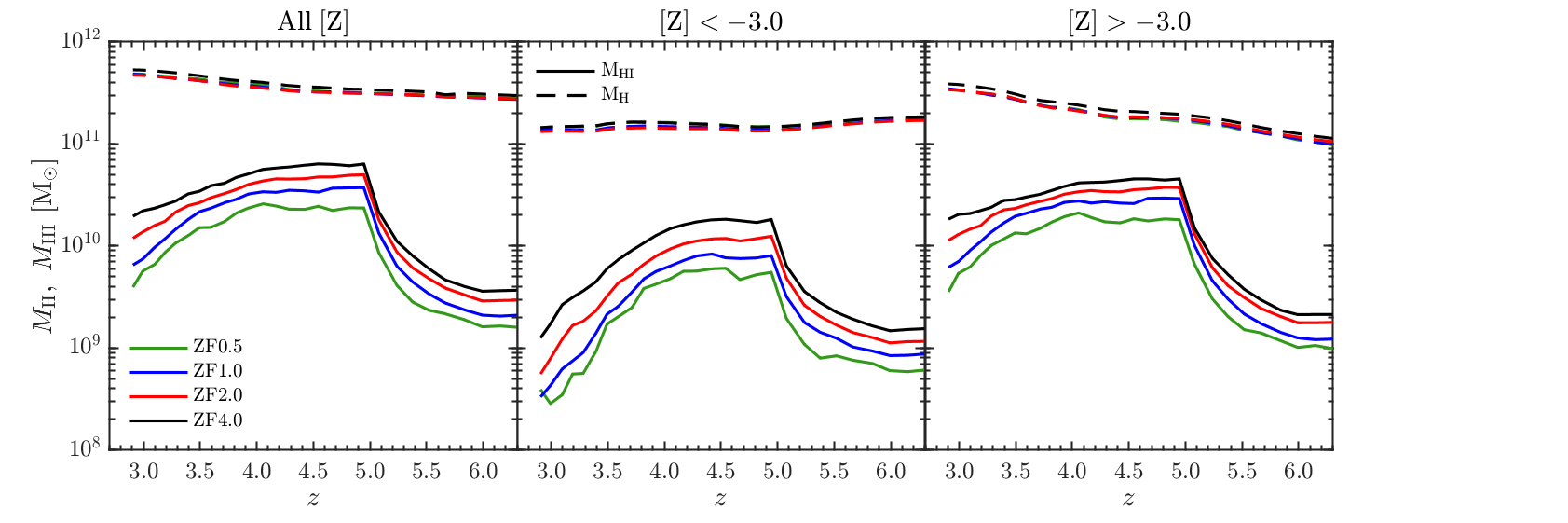}
\end{center}
\caption{Total (dashed lines) and neutral (solid lines) hydrogen mass in the IGM as a function of redshift. Masses were evaluated in the same co-moving projected area as shown in \fig{NHI_large}, namely $3.6\times 2.0~[{\rm cMpc}/h]^2$, within $\pm 100~{\rm pkpc}$ from the midplane of the sheet. All gas associated with FoF groups containing at least 32 dark matter particles was removed in order to remove the ISM and CGM. Different colours show the different resolutions, ZF0.5 in green, ZF1.0 in blue, ZF2.0 in red, and ZF4.0 in black. We show results for all gas (left), low metallicity gas representative of the IPM (centre), and high metallicity gas representative of intergalactic filaments (right). The adopted metallicity threshold is $[Z]=-3.0$ in projection in ZF1.0, as described in the text. The total hydrogen mass is converged at all times in all our simulations, in both metallicity bins. On the other hand, the HI mass systematically increases with resolution, and is $\gsim 2$ times larger in ZF4.0 than in ZF0.5 at all times. Convergence in the low metallicity gas is slightly worse, with ZF4.0 having $\gsim 3$ times more HI than ZF0.5 at all times. The rapid increase in HI mass around $z\gsim 5$ in all simulations follows the sheet collision which forms the accretion shock around the main sheet studied here.}
\label{fig:MHI}
\end{figure*}

\smallskip
Having visually identified clear differences in the morphology and HI content of intergalactic gas in filaments and the IPM as a function of simulation resolution, in this section, we aim to quantify these differences more precisely. 
Physically, we would like to do this separately for filaments and the IPM. However, a precise cell-by-cell mapping of the simulated data into different cosmic-web components using either one of the many cosmic-web finders described and compared in \citet{Libeskind18}, or the novel method inspired by the Physarum polycephalum slime mold introduced in \citet{Burchett20}, is beyond the scope of the current paper. Observationally, we would like to quantify the convergence of HI properties as a function of metallicity, since observational surveys such as KODIAQ-Z are providing data on strong HI absorbers as a function of their metal content\footnote{KODIAQ-Z are providing large samples of high-$z$ LLSs selected independent of metallicity, and then subsequently analyzing their metal content. This is contrary to some previous studies which selected strong metal-line absorbers, such as Mg-II, for follow-up spectroscopy.}. A detailed analysis of mock absorption lines produced from our highest resolution ZF4.0 simulation will be the subject of an upcoming paper (Burchett et al., in prep.). Fortunately, as evident from \fig{NHI_large}, these theoretical and observational goals are very much related, as filaments and the IPM can be roughly separated based on their metal content, with the IPM characterized by $[Z]\lsim -3.0$ and filaments by higher metallicity values. 

\smallskip
A metallicity of $\sim 10^{-3}Z_{\odot}$ also happens to be the threshold typically associated with pollution from a single PopIII supernova \citep{Wise12,Crighton16}, and is often used observationally to distinguish ``pristine'' from polluted gas. This is thus a physically meaningful threshold to distinguish filaments which are polluted by the galaxies that lie within them, from the IPM which contains hardly any star-forming galaxies. However, dividing the 3D volume of each simulation based on the metal content of individual gas cells would complicate a meaningful convergence study, since the relative volume (and mass) of the ``high-$Z$'' and ``low-$Z$'' bins would be very different in simulations with different resolutions, as evident from \fig{NHI_large}. In order to mitigate this, we hereafter assign each cell to a ``high-$Z$'' or ``low-$Z$'' bin based on whether the projected metallicity at the position of the cell in ZF1.0 (second row in \fig{NHI_large}) is above or below a threshold value of $[Z]_{\rm thresh}=-3.0$. This ensures that each metallicity bin probes the same volume in each simulation. However, this introduces a bias where cells assigned to the $[Z]<-3.0$ bin (i.e. the IPM bin) in ZF4.0 may actually have metallicity values $Z>10^{-3}Z_{\odot}$. We find that the maximal metallicity of cells assigned to the IPM in ZF4.0 is $\sim 10^{-2}Z_{\odot}$ which, for the relevant temperatures and densities, yields a cooling rate $\lsim 10\%$ larger than for pristine gas, so this should have a negligible effect on our results. Moreover, this is unrelated to the phenomena seen in \fig{nH_slice}, as all cells in this region have $Z<10^{-3}Z_{\odot}$ (see \fig{NHI_large}, region F).

\smallskip
The volume we consider is the same as in \fig{NHI_large}, namely a rectangular box with dimensions $3600~{\rm ckpc}/h\times 2000~{\rm ckpc}/h$ in the plane of the sheet, and extending $\pm 100~{\rm pkpc}$ above and below the sheet midplane. In order to focus on the IGM, we remove all galaxies and dark matter halos, including CGM gas, from our analysis. We do this by discarding all gas cells associated with FoF groups having $N_{\rm DM}\ge 32$ dark matter particles. Note that in higher resolution simulations, this removes lower mass halos and, thus, more halos in total and more total volume, offsetting somewhat the advantage afforded by using the projected metallicity in ZF1.0 to differentiate high metallicity filaments from the low metallicity IPM in all resolutions. However, this effect is negligible since the positions and volumes of halos with $\Mv\gsim 10^9\Msun$ are converged already at ZF0.5, and the volume fraction occupied by lower mass halos is negligible. We explored several other methods of removing halos, such as using a threshold in FoF mass rather than particle number, removing only gas bound to \texttt{SUBFIND} sub-halos, removing all gas within a sphere of radius $\Rv$ around all halos, and removing all gas with a 2D projected distance of $\Rv$ around all halos. All of our results and the trends with resolution are qualitatively robust to these changes, though some of the overall normalizations do change. Overall, our fiducial choice of removing all gas in FoF groups with at least 32 dark mater particles is very aggressive at removing halos, and thus conservative in terms of defining the IGM gas.

\begin{figure*}
\begin{center}
\includegraphics[trim={1.4cm 0.0cm 2.5cm 0.0cm}, clip, width =0.98 \textwidth]{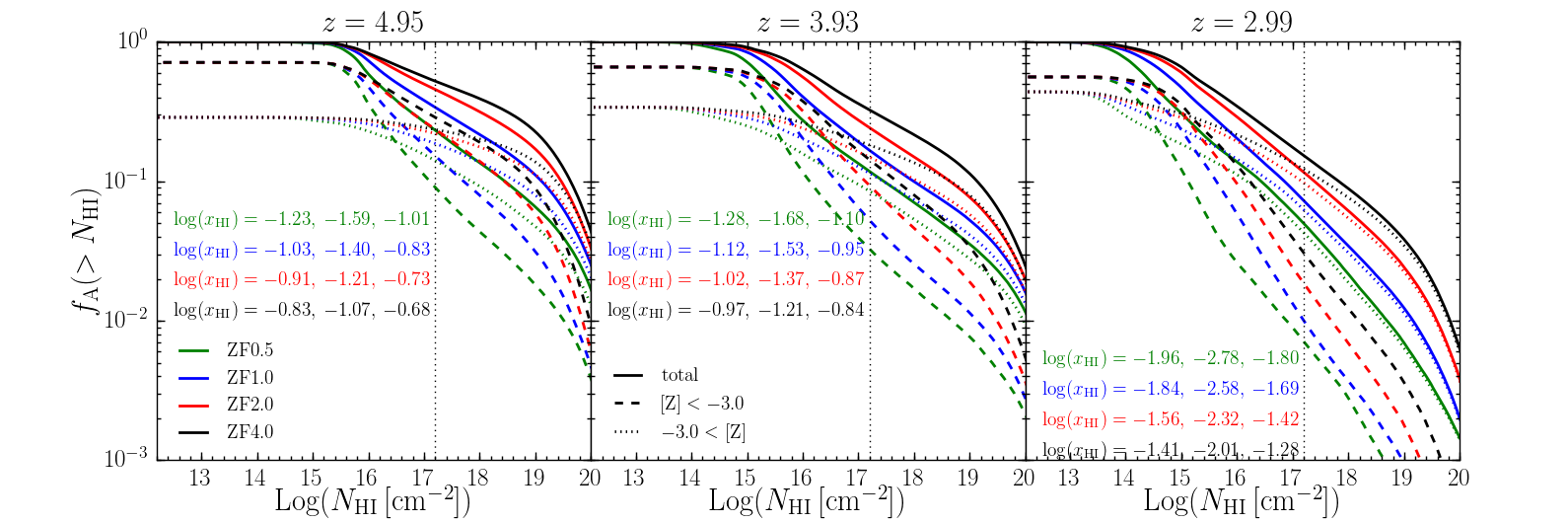}
\end{center}
\caption{Covering fractions of absorption systems within the sheet, as a function of $\NHI$. The covering fractions were computed in the same orientation as shown in \fig{NHI_large}, namely face-on with respect to the sheet, and integrated over $\pm 100 {\rm pkpc}$. When evaluating the covering fractions and $\NHI$ values, we used pixels of size $2{\rm ckpc}/h$, and removed all gas cells associated with FoF groups with $N_{\rm DM}>32$ dark matter particles. We show results at $z\sim 5$ (left), $z\sim 4$ (center), and $z\sim 3$ (right). In each panel, solid lines show the total covering fractions considering all metallicities for the absorbers, while dashed (dotted) lines show only the low (high) metallicity absorbers in the IPM (filaments), with $[Z]<-3.0$ ($[Z]>-3.0$). Different colours represent different resolutions, as in \fig{MHI}. In each panel, we list for each resolution (in the corresponding colour) the HI mass-fraction, $x_{\rm HI}=M_{\rm HI}/M_{\rm H}$, for the total, low-, and high-metallicity bins, from left to right. The HI covering fractions are not converged at any redshift in either metallicity bin. The low-metallicity gas appears less converged than the high-metallicity gas, as evidenced by the neutral fractions listed in the legend. While the $[Z]<-3.0$ absorbers occupy a larger total area at all redshifts, the highest column densities are associated with $[Z]>-3.0$ absorbers. Nonetheless, the covering fractions of LLSs with $[Z]<-3.0$ in ZF4.0 is $\sim 30\%$, $15\%$, and $3\%$ at $z\sim 5$, $4$, and $3$, while in ZF0.5 these are $\sim 9\%$, $3\%$, and $0.7\%$ respectively (see \tab{covering}). ZF4.0 displays a $\sim 1\%$ covering fraction of DLAs in the IPM, with $\NHI>10^{20}\cms$ and $Z<10^{-3}Z_{\odot}$ (see \tab{covering}). In the Ly$\alpha$ forest regime, $\NHI<10^{15}\cms$, the covering fractions are converged by ZF2.0.}
\label{fig:NHI_cover}
\end{figure*}

\smallskip
In \fig{MHI}, we show both the total hydrogen mass and the HI mass as a function of redshift, for our four resolutions. We show the results considering all gas on the left, and separately for the low-metallicity IPM (center) and the higher metallicity filaments (right). In all cases, the total hydrogen mass is well converged at all redshifts and at all resolutions. However, the HI mass systematically increases with resolution, as expected based on \fig{NHI_large}. The differences are somewhat greater in the IPM than in the filaments, with $M_{\rm HI}$ in ZF4.0 being larger than that in ZF0.5 by a factor of $\sim 3$ and $\sim 2$ in the low- and high-metallicity bins respectively, while the total HI mass is dominated by the high-metalicity bin. Likewise, $M_{\rm HI}$ in ZF4.0 is $\sim 50\%$ larger than in ZF2.0 in the low-metallicity bin, but only $\sim 25\%$ larger in the high-metallicity bin. Most importantly, there is no sign of convergence, as the HI mass fraction continues to increase with increasing resolution. The relative differences are rather constant at all redshifts, though they increase slightly towards $z\sim 3$. The rapid increase in $M_{\rm HI}$ evident in all simulations at $z\sim 5$ is due to the collision between the two smaller sheets seen in \fig{NH_large}. This collision forms the strong accretion shock visible in \fig{Temp_large}, and subsequently results in the formation of a large amount of HI. 

\smallskip
In \fig{NHI_cover}, we show the covering fractions as a function of $\NHI$ at $z\sim 5$ (left), $z\sim 4$ (center) and $z\sim 3$ (right). The covering fractions were computed face-on with respect to the sheet using pixels of $2{\rm ckpc}/h\sim 0.5$, $0.6$, and $0.7~{\rm pkpc}$ at $z\sim 5$, $4$, and $3$ respectively. This is the same pixel size used in all panels in \figs{NH_large}-\figss{nH_slice}. In each panel of \fig{NHI_cover}, we show results for all gas (solid lines), metal-poor IPM gas (dashed lines) and metal-rich filament gas (dotted lines), in each of our four resolutions (different colors). The panels also list the HI mass fraction, $x_{\rm HI}=M_{\rm HI}/M_{\rm H}$, for each metallicity bin and each resolution. These values can also be read directly from \fig{MHI}. We see that the HI covering fractions are not converged at any redshift and for either metallicity bin, especially for strong HI absorbers\footnote{The current discussion refers to the total column densities along the line of sight, with no attempt to separate individual absorbers. It is therefore possible that multiple absorbers along the same line of sight contribute to the total $\NHI$, especially at low-to-intermediate column densities. This will be addressed in an upcoming paper (Burchet et al., in prep.).} with $\NHI\gsim 10^{16}\cms$. Convergence is slightly worse in the IPM than in the filaments, as also seen in \fig{MHI}. For example, the covering fractions of LLSs with $\NHI>10^{17.2}\cms$ (vertical dotted lines in \fig{NHI_cover}) in ZF4.0 are larger than those in ZF2.0 by $\sim 50\%$ and $\lsim 10\%$ in the IPM and the filaments respectively. Likewise, they are larger than those in ZF0.5 by a factor of $\sim (4-5)$ in the IPM and $\sim 2$ in the filaments. The covering fractions of LLSs in all simulations and redshifts, in the IPM and the filaments, is presented in \tab{covering}. 
It is also interesting to note that the value of $\NHI$ where the filaments dominate over the IPM in terms of covering fraction grows larger with increasing resolution. This again implies that higher resolution drives a larger increase of $\NHI$ in the IPM than in the filaments. 

\begin{table}
\centering
\begin{tabular}{|c|c|c|c|c|c|c|c|}
\hline
\multicolumn{2}{|c|}{ } & \multicolumn{2}{c|}{LLS} & \multicolumn{2}{c|}{DLA} & \multicolumn{2}{c|}{$\mathcal{C} > 10$}\\
\hline
Sim. Name & $z$ & IPM & Fil & IPM & Fil & IPM & Fil \\
\hline
\hline
ZF0.5 & 5 & 0.09 & 0.15 & 0.004 & 0.01 & $<$0.001 & 0.002 \\
ZF1.0 & 5 & 0.15 & 0.20 & 0.005 & 0.02  & 0.007   & 0.015 \\
ZF2.0 & 5 & 0.22 & 0.22 & 0.007 & 0.03  & 0.06    & 0.06  \\
ZF4.0 & 5 & 0.30 & 0.23 & 0.01  & 0.04  & 0.23    & 0.13  \\
\hline
ZF0.5 & 4 & 0.03 & 0.09 & 0.002 & 0.01 & $<$0.001 & 0.03 \\
ZF1.0 & 4 & 0.05 & 0.12 & 0.003 & 0.01 & 0.007 & 0.015 \\
ZF2.0 & 4 & 0.09 & 0.15 & 0.004 & 0.02 & 0.06 & 0.06 \\
ZF4.0 & 4 & 0.15 & 0.18 & 0.006 & 0.02 & 0.23 & 0.13 \\
\hline
ZF0.5 & 3 & 0.007 & 0.04 & $<$0.001 & 0.001 & $<$0.001 & 0.001 \\
ZF1.0 & 3 & 0.01  & 0.06 & $<$0.001 & 0.002 & 0.003 & 0.012 \\
ZF2.0 & 3 & 0.02  & 0.10 & $<$0.001 & 0.004 & 0.03 & 0.05 \\
ZF4.0 & 3 & 0.03  & 0.12 & $<$0.001 & 0.007 & 0.10 & 0.08 \\
\hline
\end{tabular}
\caption{Covering fractions in our simulations at $z\sim 5$ (top four rows), $z\sim 4$ (middle four rows), and $z\sim 3$ (bottom four rows). The first two columns list the simulation name and redshift. Columns 3 and 4 list the covering fractions within the sheet of LLS, $\NHI>10^{17.2}\cms$, split between those located in the IPM, $[Z]<-3.0$, and the filaments, $[Z]>-3.0$ (see \fig{NHI_cover}). Columns 5 and 6 list the covering fractions within the sheet of DLAs, $\NHI>10^{20}\cms$, in the IPM and the filaments (see \fig{NHI_cover}). Columns 7 and 8 list the covering fractions within the sheet of clumping factors $\mathcal{C}=\left<n_{\rm H}^2\right>/\left<n_{\rm H}\right>^2>10$, in the IPM and filaments (see \fig{clumping}). 
}
\vspace{-13pt}
\label{tab:covering}
\end{table}

\begin{figure*}
\begin{center}
\includegraphics[trim={0.1cm 0.0cm 0.1cm 0.0cm}, clip, width =0.98 \textwidth]{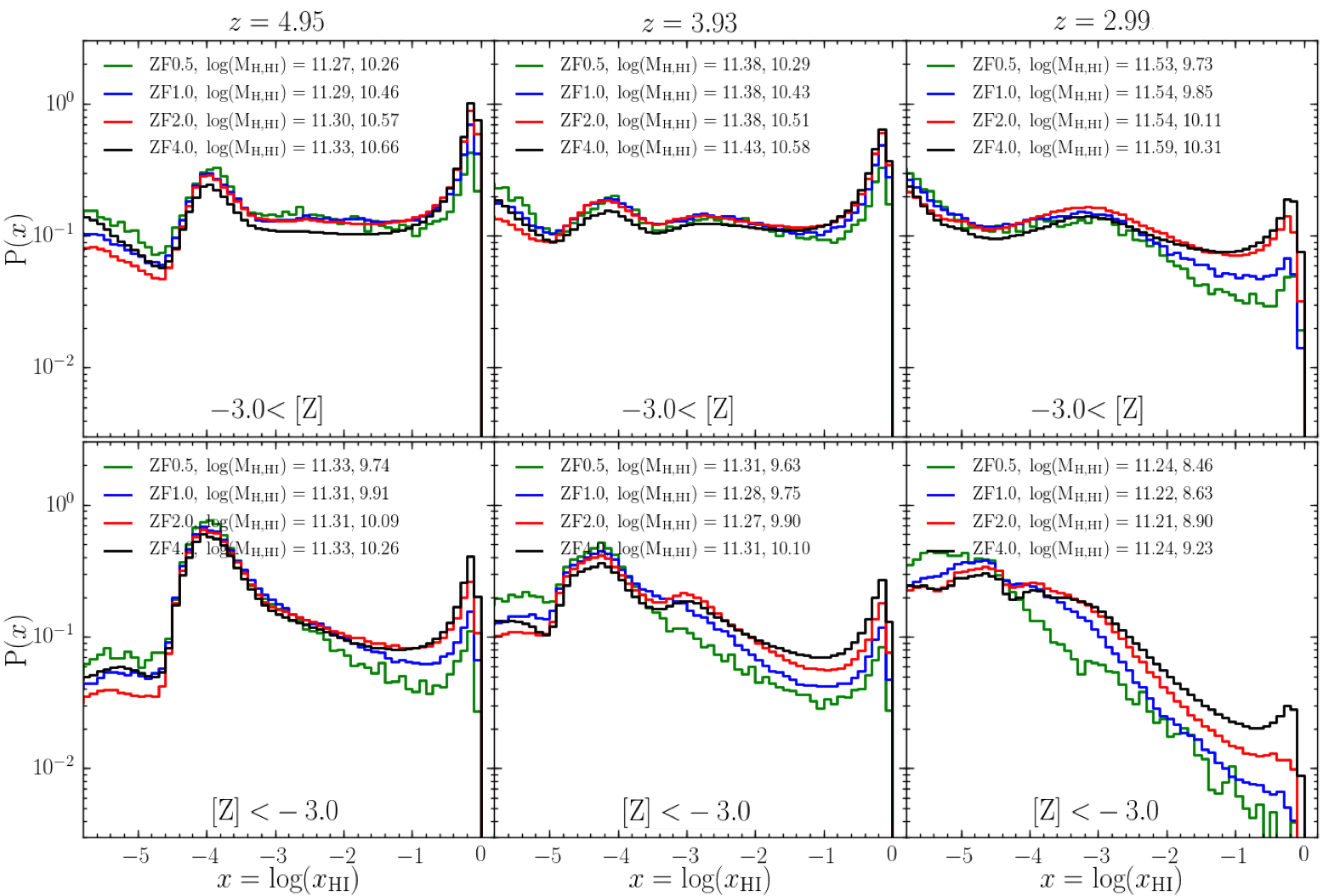}
\end{center}
\caption{Mass-weighted PDFs of neutral hydrogen fractions, $x_{\rm HI}=m_{\rm HI}/m_{\rm H}$, for gas cells within $\pm 100~{\rm pkpc}$ from the sheet midplane. As in \figs{MHI}-\figss{NHI_cover}, halos have been removed by excluding gas cells associated with FoF groups with at least 32 dark matter particles. We show results from redshift $z\sim 5$ (left), $z\sim 4$ (center), and $z\sim 3$ (right), separately for gas in the high-metallicity bin ($[Z]>-3.0$, top) and in the low-metallicity bin ($[Z]<-3.0$, bottom). Line colours refer to different resolutions, as in previous figures. Each histogram has been normalized such that the integral of $P(x)\,{\rm d}x$ over the full range of parameter space equals unity, where $x={\rm log}(x_{\rm HI})$ and the value of $P(x)$ in each bin proportional to the fraction of total gas mass in the bin. Each panel lists as well the total hydrogen mass, ${\rm M_H}$, and the neutral hydrogen mass, ${\rm M_{HI}}$, associated with each simulation resolution at each redshift and in each metallicity bin. These values can also be read from \fig{MHI}. The total hydrogen mass in the simulations is converged, while the total HI mass is not. The lack of convergence is evident from both the total $M_{\rm HI}$ values listed and the probability density near the peak in the distribution at $x_{\rm HI}\gsim 0.5$. As inferred from previous figures, the lack of convergence is worse in the low-metallicity IPM, where $M_{\rm HI}$ in ZF4.0 is typically larger by a factor $\sim 3$, $\sim 2$, and $\sim 1.5$ than $M_{\rm HI}$ in ZF0.5, ZF1.0, and ZF2.0 respectively, compared to the high-metallicity filaments, where $M_{\rm HI}$ in ZF4.0 is typically larger than in ZF0.5, ZF1.0, and ZF2.0 by factors of $\sim 2$, $\sim 1.4$, and $1.2$ respectively.
}
\label{fig:xHI_hist}
\end{figure*}

\smallskip
Also worth noting is the presence of DLAs in the IGM, with $\NHI>10^{20}\cms$. These are dominated by the filaments at all redshifts and resolutions shown (see \tab{covering}), but the covering fractions of DLAs in the IPM is $\sim 1\%$ in ZF4.0 at $z\gsim 4$. Such low metallicity DLAs are rare, but several have been observed \citep{Cooke17,Berg21}, and may potentially be evidence for IPM fragmentation as seen in our simulations. The presence of DLAs with $[Z]<-2.0$ in the IGM at $z\gsim 4$, either in filaments or the IPM, may also explain the discrepancy recently pointed out by \citet{Stern21}. By analyzing a suite of cosmological simulations, these authors found that the observed frequency of such low metallicity DLAs could not be accounted for by the ISM or the CGM of halos with $\Mv>10^9\Msun$. At the opposite end, we see that column densities associated with the Ly$\alpha$ forest, $\NHI<10^{15}\cms$, are converged to percent level at ZF2.0, in agreement with the convergence criteria of \citet{Bolton09} and \citet{Lukic15} that call for a particle mass of $m_{\rm gas}\sim 2\times 10^5\Msun$ (see \tab{sims}). While we find slight deviations at $\NHI\gsim 10^{14}\cms$ in ZF1.0 and ZF0.5 at $z\le 4$, these are minor, and their impact on the Ly$\alpha$ forest is beyond the scope of the current paper, where our focus is on denser systems. 

\begin{figure*}
\begin{center}
\includegraphics[trim={1.4cm 0.0cm 2.5cm 0.0cm}, clip, width =0.98 \textwidth]{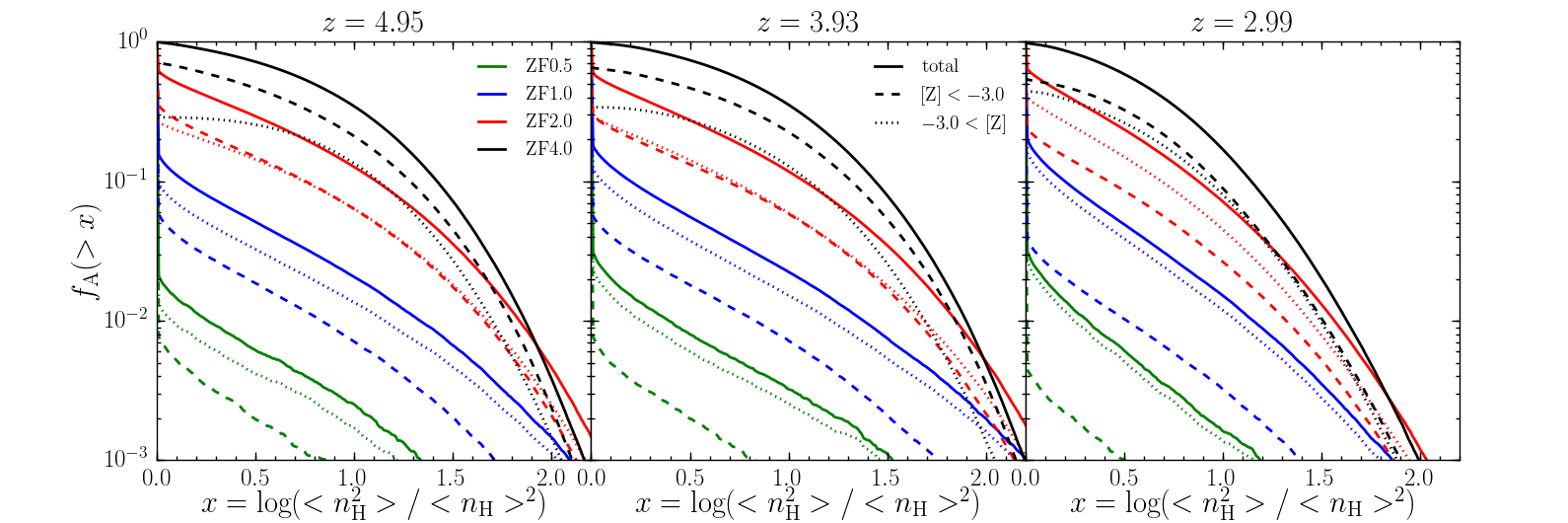}
\end{center}
\caption{Area covering fractions as a function of the line-of-sight hydrogen clumping factor, $\left<n_{\rm H}^2\right>/\left<n_{\rm H}\right>^2$. These were computed using the same frame, orientation and pixel size as the $\NHI$ covering fractions in \fig{NHI_cover}. The averages were volume-weighted, such that $\left<n_{\rm H}\right>$ is the total hydrogen mass in the column divided by the total volume of the column. As in \fig{NHI_cover}, we show results for redshift $z\sim 5$ (left), $z\sim 4$ (center), and $z\sim 3$ (right). Line colours and linestyles are the same as in \fig{NHI_cover}, with colour representing different resolutions and linestyle representing different metallicity bins. The covering fractions are not converged. In ZF0.5, $>95\%$ of the area has clumping factors $\gsim 1$ at all redshifts, while in ZF4.0, $\sim 35\%$ and $\sim 20\%$ of the sheet exhibits clumping factors $>10$ at $z\sim (4-5)$ and $z\sim 3$ respectively. In the low resolution simulations, ZF0.5 and ZF1.0, the metal rich gas exhibits larger clumping factors than the metal poor gas. For ZF2.0 these are comparable, while in ZF4.0 the low metallicity gas in the IPM is clumpier than the high-metallicity gas in the filaments.}
\label{fig:clumping}
\end{figure*}

\smallskip
In \fig{xHI_hist}, we show the mass-weighted PDF of neutral mass fractions, $x_{\rm HI}=m_{\rm HI}/m_{\rm H}$, among all IGM gas cells in the simulations, in 3D (not in projection). We show results at $z\sim 5$ (left), $z\sim 4$ (center), and $z\sim 3$ (right), separately for the metal-rich filaments (top) and the metal-poor IPM (bottom). In each panel, we list the total hydrogen mass and HI mass in the corresponding metallicity bin for each resolution. These numbers can also be read from \fig{MHI}, and show that while $M_{\rm H}$ is converged, $M_{\rm HI}$ is not. At all redshifts and in each metallicity bin, the distribution of $x_{\rm HI}$ values in the ZF4.0 simulation has a peak at high neutral fractions, $x_{\rm HI}\sim (0.5-0.6)$, and the strength of this peak decreases as the simulation resolution is decreased. The relative strength of the peak and the level of convergence with resolution are both very similar at $z\sim 5$ and $z\sim 4$. However, at $z\sim 3$, the relative contribution of this peak declines significantly at all resolutions, and convergence appears worse, as also evident from the values of $M_{\rm HI}$. The decline in neutral fractions at $z\sim 3$ is due both to the stronger UV background, including additional photoheating by AGN, and the overall lower densities which reduce the amount of self-shielding. As in \figs{MHI} and \figss{NHI_cover}, convergence is much better in the high-metallicity filaments than the low-metallicity IPM. For gas with $[Z]>-3.0$, the probability density near the high-$x_{\rm HI}$ peak in ZF4.0 is nearly identical to ZF2.0 at $z\sim 5$ and $4$ (though ZF1.0 and ZF0.5 are both noticeably smaller), and is $\sim 30\%$ larger at $z\sim 3$. On the other hand, for gas with $[Z]<-3.0$, the probability densities in ZF4.0 are larger than in ZF2.0 by $\sim 50\%$ at $z\sim 5$ and $4$ and by a factor $\sim 3$ at $z\sim 3$.

\smallskip
In \fig{clumping}, we show the covering fractions of the gas clumping factor, defined as $\mathcal{C}\equiv \left<n_{\rm H}^2\right>/\left<n_{\rm H}\right>^2$, where $n_{\rm H}$ is the volume density of total hydrogen, and $\left< \cdot \right>$ denotes a volume weighted average along the line-of-sight. This quantity is very important for evaluating IGM properties, such as the optical depth or ionization state \citep[e.g.][and references therein]{Pawlik09}. Large clumping factors are also a natural outcome of shattering and a good metric of thermal instability in a gaseous medium \citep{McCourt18}. We show results for $z\sim 5$ (left), $z\sim 4$ (center), and $z\sim 3$ (right). In each panel, we show results for all gas, metal-poor IPM, and metal-rich filaments using solid, dashed, and dotted lines respectively. As expected from the visual impression in \fig{nH_slice}, the clumping factor is far from converged, at all redshifts and in each metallicity bin. Consistent with previous results, we find the convergence is worse in the IPM than in the filaments. This is most evident here by noting that in ZF4.0, the covering fractions in the IPM are larger than in the filaments for all values of $\mathcal{C}>1$, while in ZF2.0, ZF1.0, and ZF0.5 the covering fractions are larger in the filaments. The IPM is thus significantly clumpier in ZF4.0 than in lower resolution simulations, while there is a smaller difference in the degree of clumpiness in filaments. The covering fraction of sightlines with $\mathcal{C}>10$ in the ZF4.0 IPM is $\sim 23\%$ at $z\sim 5$ and $4$, and $\sim 10\%$ at $z\sim 3$ (see \tab{covering}). Examining the covering fractions for total gas (solid lines), we see that only in ZF4.0 is this curve continuous as the clumping factor approaches unity from above. In lower resolution simulations, a large fraction of the area has $\mathcal{C}\sim 1$, $\sim 40\%$, $\sim 80\%$, and $\sim 97\%$ in ZF2.0, ZF1.0, and ZF0.5 respectively, leading to a sharp jump in the covering fractions as the clumping factor approaches unity\footnote{By definition, $\left<n_{\rm H}^2\right>/\left<n_{\rm H}\right>^2 ~\ge 1$, so the covering fraction of clumping factors greater than or equal to 1 must be unity.}.

\section{Gas Thermal Properties in the IPM} 
\label{sec:results_3}

\smallskip
In the previous section, we showed that the HI masses, column densities, and clumpiness in the IGM are not converged in our simulations, and that convergence was worse in the metal-poor IPM than in the more metal-rich filaments. We discuss potential reasons for the better convergence in filaments in \se{disc}. Here, we focus on the metal-poor IPM and examine the thermal properties of the gas, to try and understand what might be leading to the lack of convergence in HI properties. All of the trends we report in this section are also found in the filament gas, but to a lesser degree, reflective of the slightly better convergence in HI properties. 

\begin{figure*}
\begin{center}
\includegraphics[trim={0.2cm 0.0cm 0.0cm 0.0cm}, clip, width =0.60 \textwidth]{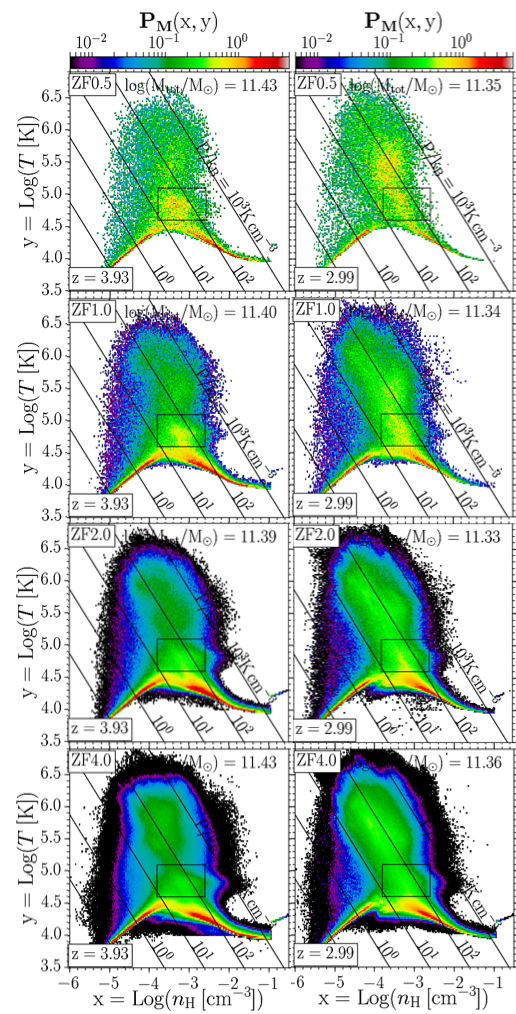}
\end{center}
\vspace{-10.0pt}
\caption{
Phase diagrams for $[Z]<-3.0$ gas within $\pm 100~{\rm pkpc}$ from the sheet midplane at $z\sim 4$ (left) and $z\sim 3$ (right), after removing halos. Colour represents a normalized probability density, such that the integral of ${\bf{P}}(x,y)\,{\rm d}x\,{\rm d}y$ over the full range equals unity, where $x={\rm log}(n_{\rm H})$ and $y={\rm log}(T)$ and the value of ${\bf{P}}(x,y)$ in each bin is proportional to the gas mass fraction in the bin. Different panels show the different simulation resolutions, from ZF0.5 (top) to ZF4.0 (bottom). Each panel also lists the total gas mass included in the calculation, which is converged across all simulations at each redshift. Diagonal lines show constant thermal pressure, from $P/k_{\rm B}=0.1-1000~{\rm K~cm^{-3}}$, as marked. 
Low pressure gas with $P/k_{\rm B} < 10~{\rm K~cm^{-3}}$ represents the diffuse IGM outside the sheet, which has not been shock-heated at the sheet boundary. 
The black rectangle spanning $4.6<{\rm log}(T/{\rm K})<5.1$ and $-3.8<{\rm log}(n_{\rm H}/\cmc)<-2.6$ highlights a region of phase-space where the gas mass-fraction seems to systematically decrease as the resolution is increased, creating an ``overdensity'' in the low-resolution simulations. At $z\sim 4$ this overdensity seems roughly isothermal and is more prominent than at $z\sim 3$ where it seems roughly isobaric. 
}
\label{fig:phase_low}
\end{figure*}

\smallskip
In \fig{phase_low}, we show the mass-weighted distribution of IPM gas in density-temperature space for our four resolutions, at $z\sim 4$ (left) and $z\sim 3$ (right). The total gas mass (rather than hydrogen mass) represented by these distributions is listed in each panel, and is consistent to better than $10\%$ at all resolutions. Several features are apparent in these distributions. The ridge-line at low densities and temperatures, $T\lsim 10^{4.5}\K$ and $n_{\rm H}\lsim 10^{-3.8}\cmc$, represents pre-shock gas above or below the accretion shock surrounding the sheet but within $\pm 100 {\rm pkpc}$ from the midplane. The ridge-line thus shows the mean $T-\rho$ relation of the diffuse IGM, $T\propto \rho^{\gamma-1}$, with $\gamma\sim 1.5$, consistent with previous estimates at $z\sim 4$ \citep[e.g.][]{Lukic15}. The ridge-line at higher densities, $T\lsim 10^{4.5}\K$ and $n_{\rm H}\gsim 10^{-2.8}\cmc$, represents IPM gas in thermal equlibrium with the UVB. The narrow sliver of gas with $n_{\rm H}>0.13\cmc$, more prominent at higher resolution, represents the artificial equation of state  implemented in the simulation for star-forming gas \citep{Springel03}. While it is fascinating that some star-formation may occur in the IPM, far from any galaxy or halo resolved by $>32$ dark matter particles, this is very likely a product of the simplified star-formation recipe implemented in the simulation. Furthermore, even in ZF4.0 this represents a negligible fraction of the total IPM mass. The resulting SFR is thus not expected to influence the IPM overall, and we do not focus on this further. 

\smallskip
At $z\sim 4$, most of the gas with $T>10^5\K$ in ZF4.0 and ZF2.0 is roughly isobaric with thermal pressure $P/k_{\rm B}\sim 100\K\,\cmc$. However, in ZF1.0 and ZF0.5, gas with $T<10^6\K$ is distributed more isochorically. This is most evident when looking at the ridge line at $T\sim (10^5-10^{5.3})\K$ and $P/k_{\rm B}\lsim 100\K\,\cmc$. But the most prominent difference in the distribution at different resolutions is the excess probability density at $T\sim (10^{4.6}-10^{4.9})\K$ and $n_{\rm H}\sim (10^{-3.6}-10^{-3.0})\cmc$ in low resolution simulations, highlighted by a black box in each panel of \fig{phase_low}. A similar feature was noted in the CGM simulations of \citet{Hummels19} (see their figure 6), though there the low resolution simulation had a large excess of gas at higher temperatures, $T\gsim 10^{5.5}\K$, with only a modest excess of gas in the density and temperature range we are discussing here. 

\begin{figure*}
\begin{center}
\includegraphics[trim={1.4cm 0.0cm 2.5cm 0.0cm}, clip, width =0.98 \textwidth]{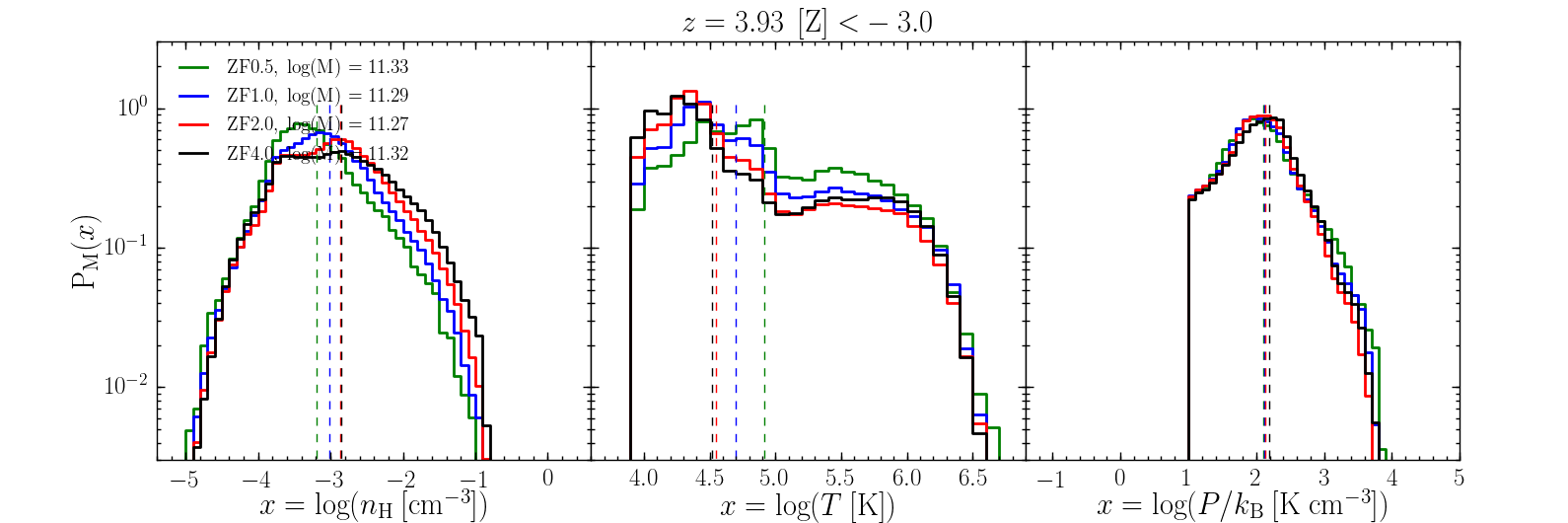}
\end{center}
\caption{Mass-weighted PDFs of density (left), temperature (center), and thermal pressure (right), for gas with $[Z]<-3.0$ within $\pm 100~{\rm pkpc}$ from the sheet midplane at $z=4$, after removing halos. To focus on IPM gas, we only consider gas with pressure $P/k_{\rm B}>10~\K\,\cmc$ (see \fig{phase_low}). Different colours represent the different simulation resolutions, and the vertical dashed lines show the medians. Each curve represents a normalized probability density, such that the integral of ${\rm P_M}(x)\,{\rm d}x$ over the full range of parameter space equals unity, where $x={\rm log}(n_{\rm H})$ (left), ${\rm log}(T)$ (center), or ${\rm log}(P/k_{\rm B})$ (right). The total gas mass represented by each PDF is listed in the legend, and is very similar among all resolutions. The pressure distribution is also well converged at all resolutions. However, the fraction of gas mass at intermediate temperatures and densities, $4.6\lsim {\rm log}(T/{\rm K})\lsim 5.0$ and $-3.8\lsim {\rm log}(n_{\rm H}/{\cmc})\lsim -3.0$, systematically increases as the resolution is decreased, as evident from the region highlighted by the black rectangle in \fig{phase_low}. At the same time, the fraction of gas mass at low temperatures and high densities, ${\rm log}(T/{\rm K})\lsim 4.5$ and ${\rm log}(n_{\rm H}/{\rm \cmc})\gsim -2.4$, systematically increases as the resolution is \textit{increased}. This is suggestive of a ``cooling bottleneck'', where gas ``piles-up'' at $T\lsim 10^5\K$ in low resolution simulations and cannot cool to temperatures $T\lsim 10^{4.5}\K$.
}
\label{fig:dens_low}
\end{figure*}


\smallskip
In \fig{dens_low}, we show the mass-weighted PDFs of density, temperature, and pressure for IPM gas at $z\sim 4$. In order to focus only on post-shock gas actually within the sheet, and remove the low-temperature, low-density pre-shock IGM above or below the sheet, we only consider here gas with pressures $P/k_{\rm B}>10\K\,\cmc$, which explains the sharp cutoff in the pressure distribution. The pressure PDFs are very similar across all resolutions. This is sensible, as the gas pressure is determined by the ram pressure of infalling gas onto the sheet, which forms the large-scale accretion shock sandwiching the sheet. The density of this infalling gas is roughly the Universal mean baryon density, while its velocity is set by the gravitational acceleration of the sheet itself, which is converged at all resolutions. In the temperature PDFs, several interesting features are apparent. At low temperatures, $T<10^{4.5}\K$, the gas mass fraction increases monotonically with resolution, and is not yet converged. This is related to the lack of convergence of HI mass, as nearly all of the HI is in this temperature range. This excess of cold gas at high resolution is offset by a deficiency of warm gas, with $T\sim (10^{4.6}-10^5)\K$, the same temperature range where we saw the enhanced probability density in low resolution simulations in \fig{phase_low}. Comparing these two temperature ranges, it seems as though there is a ``cooling bottleneck'' preventing gas in low resolution simulations from cooling below $\sim 10^5\K$, and causing gas to ``pile-up'' at these temperatures. We examine this further below, and discuss potential physical explanations for this in \se{disc}. At $T\sim (10^5-10^6)\K$, the same temperature range where \citet{Hummels19} found a large excess in gas mass in their low resolution CGM simulations, ZF0.5 exhibits excess mass compared to higher resolution simulations. However, unlike the pile-up at $T\sim (10^{4.6}-10^5)\K$, this does not seem to be monotonic with resolution. ZF4.0 has more mass than ZF2.0 in this regime, and is very similar to ZF1.0. The density PDF displays a monotonic excess of mass in high resolution simulations at $n\gsim 10^{-2}\cmc$, and a monotonic excess of mass in low resolution simulations at $n\sim (10^{-4}-10^{-3})\cmc$. Given the similar pressure distributions, these correspond to the trends in the temperature PDF at $T<10^{4.5}\K$ and $T\sim (10^{4.6}-10^5)\cmc$, respectively. 

\begin{figure*}
\begin{center}
\includegraphics[trim={1.4cm 0.0cm 2.5cm 0.0cm}, clip, width =0.98 \textwidth]{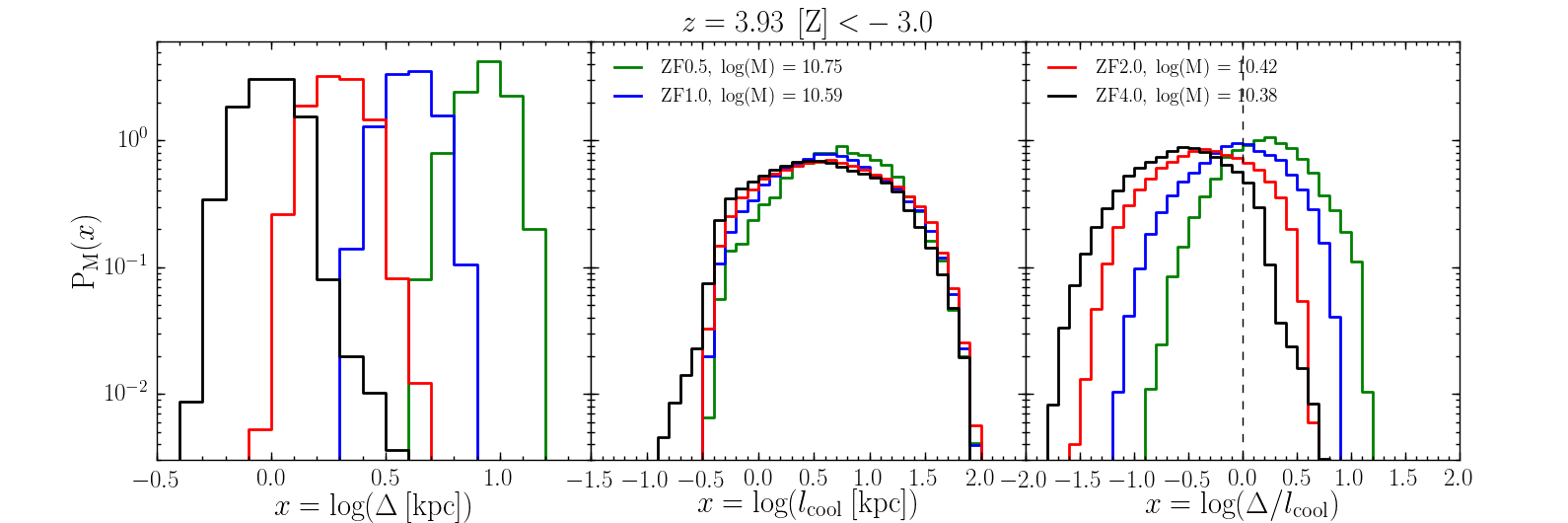}
\end{center}
\caption{Mass-weighted PDFs of cell-size, $\Delta$ (left), cooling length, $l_{\rm cool}=\cs t_{\rm cool}$ (center), and the ratio $\Delta/l_{\rm cool}$ (right, with the vertical dashed line marking a ratio of unity), for gas with $[Z]<-3.0$ within $\pm 100~{\rm pkpc}$ from the sheet midplane at $z=4$, after removing halos, with densities and temperatures in the range marked in \fig{phase_low}, $4.6<{\rm log}(T/{\rm K})<5.1$ and $-3.8<{\rm log}(n_{\rm H}/\cmc)<-2.6$. Different colours represent the different simulation resolutions. Each curve represents a normalized probability density, such that the integral of ${\rm P_M}(x)\,{\rm d}x$ over the full range of parameter space equals unity, where $x={\rm log}(\Delta)$ (left), ${\rm log}(l_{\rm cool})$ (center), or ${\rm log}(\Delta/l_{\rm cool})$ (right). The total gas mass represented by each PDF is listed in the legend, and decreases with increasing resolution, as deduced from \figs{phase_low}-\figss{dens_low}. In ZF0.5, there is roughly 2.5 times more gas in this range of temperatures and densities than in ZF4.0. The distribution of cooling lengths is very similar among all simulation resolutions, except a small tail towards low values of $l_{\rm cool}$ in ZF4.0. Since the sound speed in this temperature range spans a factor of $\sim 1.7$, this implies that the cooling times are well converged. In ZF2.0 and ZF4.0, the cooling length is resolved in most of the gas, with the median ratio of $\Delta/l_{\rm cool}<1$. However, in ZF1.0 and ZF0.5, the cooling length is typically unresolved, with the median ratio of $\Delta/l_{\rm cool}>1$. 
}
\label{fig:cell_low}
\end{figure*}

\smallskip
In \fig{cell_low}, we focus on gas in the region of temperature-density space highlighted by the black rectangle in \fig{phase_low}, $4.6 \le {\rm log}(T/\K)\le 5.1$ and $-3.8 \le {\rm log}(n_{\rm H}/\cmc)\le -2.6$, where there appears to be a ``pile-up'' in low resolution simulations. The total gas mass in this region is listed in the figure legend for each resolution, and we see that indeed the mass in this region in ZF2.0, ZF1.0, and ZF0.5 is $\sim 0.04$, $0.21$, and $0.37$ dex larger than in ZF4.0. We show mass-weighted PDFs of the cell sizes $\Delta={\rm vol}^{1/3}$ with ${\rm vol}$ the cell volume (left), the cooling length $l_{\rm cool}=c_{\rm s}t_{\rm cool}$ (center), and the ratio of cell size to cooling length (right). Note that this $l_{\rm cool}$ is \textit{not} the minimal value reached at $T\gsim 10^4\K$ \citep{McCourt18}, but rather the local cooling length at the current temperature and density of the gas. The typical cell size increases by a factor of $\sim 2$ between each resolution level, as expected, from $\sim 1\kpc$ in ZF4.0 to $\sim 8\kpc$ in ZF0.5. The distributions of cooling lengths, on the other hand, are very similar at all resolutions, save for a small tail in ZF4.0 towards very small $l_{\rm cool}$ which contains $\lsim 0.5\%$ of the mass. The typical cooling length is $l_{\rm cool}\sim 3\kpc$, which given the typical temperature of $T\sim 10^{4.8}\K$ and corresponding sound speed of $c_{\rm s}\sim 38\kms$, yields a cooling time of $t_{\rm cool} \lsim 100\Myr$. Examining the PDF of $\Delta/l_{\rm cool}$, we see that in ZF4.0 and ZF2.0 the cooling length is at least marginally resolved for most of the gas mass, while in ZF1.0 and ZF0.5 it is not. A cell that is larger than $l_{\rm cool}$ cannot cool isobarically. Instead, it must either cool isochorically, or else ``shatter'' into fragments of size $l_{\rm cool}$ which proceed to cool isobarically. However, if $l_{\rm cool}$ is unresolved in a simulation, the latter path closes, and the gas cell must resort to isochoric cooling. This is almost certainly the reason why the ridge-line connecting this region to lower temperatures in the phase diagrams of \fig{phase_low} seem to transition from mostly isochoric in ZF0.5 to mostly isobaric in ZF4.0. 

\smallskip
The results at $z\sim 5$ are extremely similar to $z\sim 4$ and are not shown here. The same mass excess in low resolution simulations at $T\sim (10^{4.6}-10^5)\K$ is present, along with a comparable mass excess in high resolution simulations at $T\lsim 10^{4.5}\K$. This gives the same impression of a ``cooling bottleneck'' as discussed above. 
The distribution of $l_{\rm cool}$ values in the same region of $nT$ space studied in \fig{cell_low} is even better converged than at $z\sim 4$, with a characteristic cooling time of $t_{\rm cool}\sim 50\Myr$. $l_{\rm cool}$ is resolved in most of the gas mass in ZF2.0 and ZF4.0, but unresolved in ZF1.0 and ZF0.5.

\smallskip
On the other hand, at $z\sim 3$, the excess probability density in this region is less pronounced than at $z\sim 4$, as can be seen in the right-hand column of \Fig{phase_low}. Moreover, to the extent that there is an overdensity in this region at $z\sim 3$, it reflects an isobaric distribution rather than the roughly isothermal distribution of the overdensity at $z\sim 4$. Similarly, the overall distribution in $nT$ space in ZF1.0 and ZF0.5 seems more isobaric at $z\sim 3$ than it did at $z\sim 4$. On the other hand, the excess probability density at $T\sim 10^{5.5}\K$ in low resolution simulations appears more prominant at $z\sim 3$ than it did at $z\sim 4$. This is a very similar feature to that found by \citet{Hummels19} in their CGM simulations at $z\sim 1$. While we do not show here PDFs of the thermal properties at $z\sim 3$, we report that the excess mass at $T\sim (10^{4.6}-10^5)\K$ in low resolution simulations seen in \fig{dens_low} has all but disappeared by $z\sim 3$, while the gas mass with $T<10^{4.5}$ shows much better convergence than it did at $z\sim 4$, beginning with ZF1.0. The distribution of $l_{\rm cool}$ values is reasonably well converged, with a characteristic cooling time of $t_{\rm cool}\sim 200\Myr$. Unlike at $z\sim 4$, $l_{\rm cool}$ is resolved for most of the mass in ZF1.0 and even in ZF0.5 at $z\sim 3$. This suggests a correlation between the ``cooling bottleneck'' at $T\lsim 10^5\K$ and the fraction of mass at these temperatures where $l_{\rm cool}$ is resolved.

\begin{figure*}
\begin{center}
\includegraphics[trim={0.2cm 0.0cm 0.1cm 0.0cm}, clip, width =0.98 \textwidth]{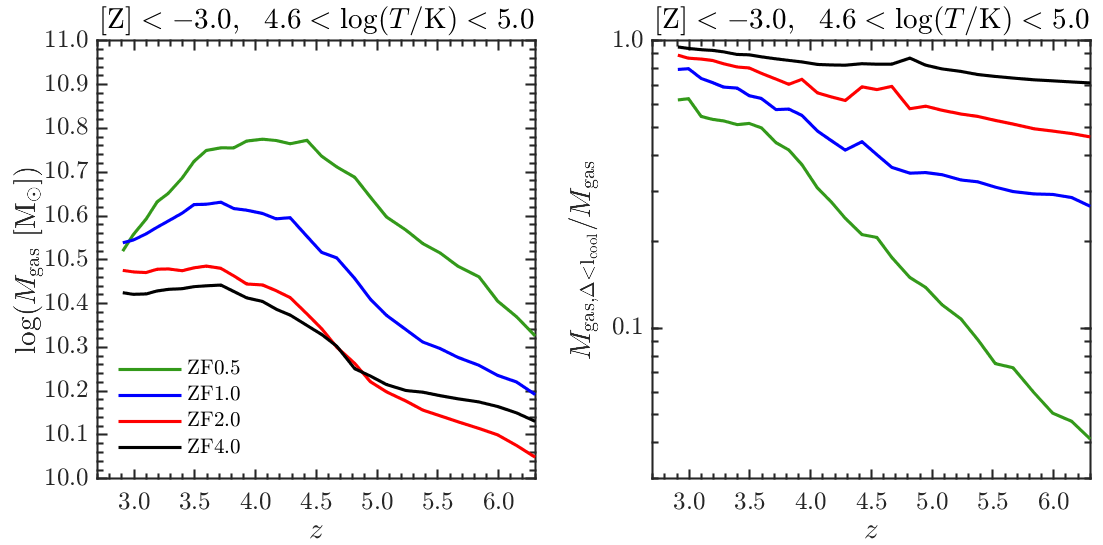}
\end{center}
\caption{The mass of gas with temperatures $4.6<{\rm log}(T/\K)<5.0$ (left), and the mass fraction of cells where the cooling length is resolved (right), as a function of redshift. Different colours represent different simulation resolutions, as indicated. In all simulations, the mass fraction of cells where the cooling length is resolved increases with decreasing redshift, due to declining densities and pressures resulting in larger values of $l_{\rm cool}$. At $z\sim 5$, ZF0.5 has $\sim 2.5$ times more gas than ZF4.0 in this temperature range, while the mass fraction of gas resolving $l_{\rm cool}$ decreases from $\sim 80\%$ in ZF4.0 to $\sim 13\%$ in ZF0.5. At $z\sim 3$, the corresponding fractions are $\sim 95\%$ and $\sim 60\%$, while the mass of gas in this temperature range is only $\sim 40\%$ larger in ZF0.5 than in ZF4.0.
}
\label{fig:Mwarm}
\end{figure*}

\smallskip
In \fig{Mwarm}, we show further evidence for such a correlation. On the left, we show as a function of redshift the mass of gas in the IPM, i.e., with metallicity ${Z}<-3.0$ and thermal pressure $P/k_{\rm B}>10\K\,\cmc$, in the temperature range $10^{4.6}<T/\K < 10^{5}$. In all simulations, this mass increases following the sheet collision at $z\sim 5$. However, the mass increase following the collision grows smaller as the resolution increases, from ZF0.5 to ZF4.0. Prior to the shock, at $z>5$, ZF4.0 has slightly more mass in this temperature range than ZF2.0, though we suspect this is a consequence of more star-formation in lower mass galaxies at earlier times in the higher resolution simulation, as inferred from the metallicity distribution (\fig{NHI_large}). However, the gas mass in these two simulations is within $\lsim 0.05$ dex at all times. On the other hand, the excess mass in ZF1.0 and ZF0.5 compared to ZF4.0 increases following the shock, peaking at $z\sim 4.5$ at $\sim 0.2$ and $\sim 0.4$ dex respectively. At later times, the warm gas mass in these simulations noticeably declines, contrary to ZF2.0 and ZF4.0 where it remains roughly constant. By $z\sim 3$ the mass excess in low resolution simulations compared to ZF4.0 is $\lsim 0.15$ dex, or $\sim 40\%$. 

\smallskip
In the right-hand panel of \fig{Mwarm}, we show as a function of redshift the fraction of the gas mass in the left-hand panel for which $l_{\rm cool}$ is resolved. In all simulations, this fraction increases monotonically with time as the Universe expands and the characteristic densities and pressures decrease, while simultaneously the UVB increases thus lowering the cooling rates. The decline in warm gas mass in ZF1.0 and ZF0.5, which begins around $z\gsim 3.5$, coincides with the time where the mass fraction of resolved $l_{\rm cool}$ is comparable to the fraction in ZF2.0 following the shock formation at $z\sim 5$. Overall, towards $z\sim 3$ as the simulations converge in terms of the fraction of mass where $l_{\rm cool}$ is resolved, the total gas mass in this temperature range also begins to converge.

\section{The Origin of Enhanced Cooling in High-Resolution Simulations} 
\label{sec:disc}


\smallskip
In the previous section, 
we found that the mass of cold gas with $T\lsim 10^{4.5}\K$ in the IPM systematically decreases in lower resolution simulations, while the mass of warm gas with $T\sim (10^{4.6}-10^{5})\K$ systematically increases. Qualitatively, this seemed to suggest the presence of a ``cooling bottleneck'' in low resolution simulations, where the cooling efficiency at $T\lsim 10^{5}\K$ is reduced causing gas to ``pile-up'' at these temperatures. We further saw that this excess mass in low-resolution simulations correlates with the fraction of mass at these temperatures where the cooling length is resolved. 
In this section, we begin in \se{disc_phys} by discussing the physical significance of resolving the cooling length at $T\lsim 10^5\K$, and how this can affect gas cooling to lower temperatures. In \se{disc_numer} we address additional physical and numerical effects which can affect the formation of cold, neutral gas in the simulations. Finally, in \se{disc_Fil_IPM}, we speculate as to why intergalactic filaments are better converged than the IPM.

\subsection{The physical importance of resolving $l_{\rm cool}$ at $T\lsim 10^5\K$}
\label{sec:disc_phys}

\smallskip
Since its importance as a characteristic scale for cold gas clouds in a multiphase medium was highlighted in \citet{McCourt18}, a lot of emphasis has been placed on resolving the minimal cooling length at $T\gsim 10^4\K$, $l_{\rm cool,min}\sim 100\pc\,(n/10^{-3}\cmc)^{-1}$, in order to achieve converged results in simulations. However, while this scale remains at best marginally resolved in ZF4.0, and totally unresolved in all other simulations, we find a correlation between the presence of cold and neutral gas and whether or not the cooling length at $T\lsim 10^5\K$ is resolved. We propose below two potential explanations for this correlation, highlighting the physical significance of this scale, and why not resolving it can significantly hinder the formation of cold and neutral gas in the IPM, and by extension in the IGM and CGM in general. 

\subsubsection{Isochoric Thermal Stability Coupled with Unresolved Isobaric Thermal Instability}
\label{sec:iso_stab}

\smallskip
The process of shattering, as described by \citet{McCourt18}, occurs when cooling clouds are larger than the cooling length at their current temperature, or equivalently, when the cooling time becomes shorter than the sound crossing time in the cloud. Such clouds cannot maintain sonic contact while cooling, and they thus shatter into smaller fragments that proceed to cool isobarically, maintaining pressure equilibrium with their surroundings. This is seemingly contrary to previous studies which assumed that such clouds would simply cool isochorically, i.e. at constant density, and regain pressure equilibrium at a later time by rapid contraction after the cloud had cooled \citep[e.g.][]{Burkert00}. A resolution to this apparent contradiction was recently proposed by \citet{Das21} (see also \citealp{Waters19} for an alternate perspective). These authors suggested that whether or not a large cloud, with size $R_{\rm c}>l_{\rm cool}$, will shatter depends on whether or not the isochoric mode of thermal instability is unstable at the initial cloud temperature. 

\smallskip
To elaborate on this slightly, when one conducts a linear analysis of thermal instability, one must distinguish between isochoric modes, where the density of the initial perturbation remains constant, and isobaric modes, where the pressure of the initial perturbation remains constant. While a given perturbation does not have to be perfectly isochoric or isobaric, these represent two limits of the instability \citep[e.g.][]{Field65,Burkert00,Das21}. Isobaric modes always grow faster than isochoric modes, because the cooling rate is proportional to the density squared, ${\dot{E}}\propto n^2\Lambda(T)$, so the increase in density as isobaric modes cool enhances the cooling rate. These two modes not only have different growth rates when they are unstable, but they have different conditions for instability. In general, at a given temperature, isobaric and isochoric modes can both be unstable, they can both be stable, or isobaric modes can be unstable while isochoric modes are stable. Now assume an initially large cloud, with $R_{\rm c}>l_{\rm cool}$, starting from near thermal equilibrium while pressure confined by an external medium. If the initial conditions in the cloud are unstable to isochoric thermal instability, the cloud will cool isochorically\footnote{While small-scale isobaric modes can still grow faster than the large-scale isochoric mode, these coalesce as the large-scale isochoric mode cools, preventing the formation of a shattered structure.} 
and regain pressure equilibrium at the end stage of cooling \citep{Das21}. On the other hand, if the initial conditions of the cloud are such that isochoric modes are stable while isobaric modes are unstable, the cloud will be unable to monolithically cool. Rather, isobaric perturbations on small scales, $\Delta<l_{\rm cool}$, will grow causing the cloud to ``shatter'' and resulting in a mist of small cold cloudlets at the end of the cooling process \citep{Das21}. 

\smallskip
\citet{Das21} confirmed their model using a series of 1D simulations of cooling clouds, where they were always able to resolve the initial cooling lengths. However, this begs the questions what might happen in a simulation where isochoric modes are stable in the initial cloud, but the initial cooling length is unresolved. In such a scenario, the cloud cannot cool isochorically, yet it also cannot cool isobarically since this can only happen on scales $\Delta<l_{\rm cool}$ which are unresolved. We posit that this will result in a cooling bottleneck where linear modes are stable and only non-linear modes will be able to cool, and we speculate that this is the reason for the correlation between the excess gas mass in the IPM with $T\lsim 10^5\K$ and the fraction of mass at this temperature where $l_{\rm cool}$ is resolved. 
There are two necessary conditions for this hypothesis to be valid: (1) there exists an approximate thermal equlibrium near $T\sim 10^5\K$ in the IPM, and (2) this is in a regime where isochoric modes are stable while isobaric modes are unstable. We demonstrate both of these conditions in Appendix \se{app}.

\begin{figure*}
\begin{center}
\includegraphics[trim={0.6cm 0.18cm 1.4cm 0.6cm}, clip, width =0.98 \textwidth]{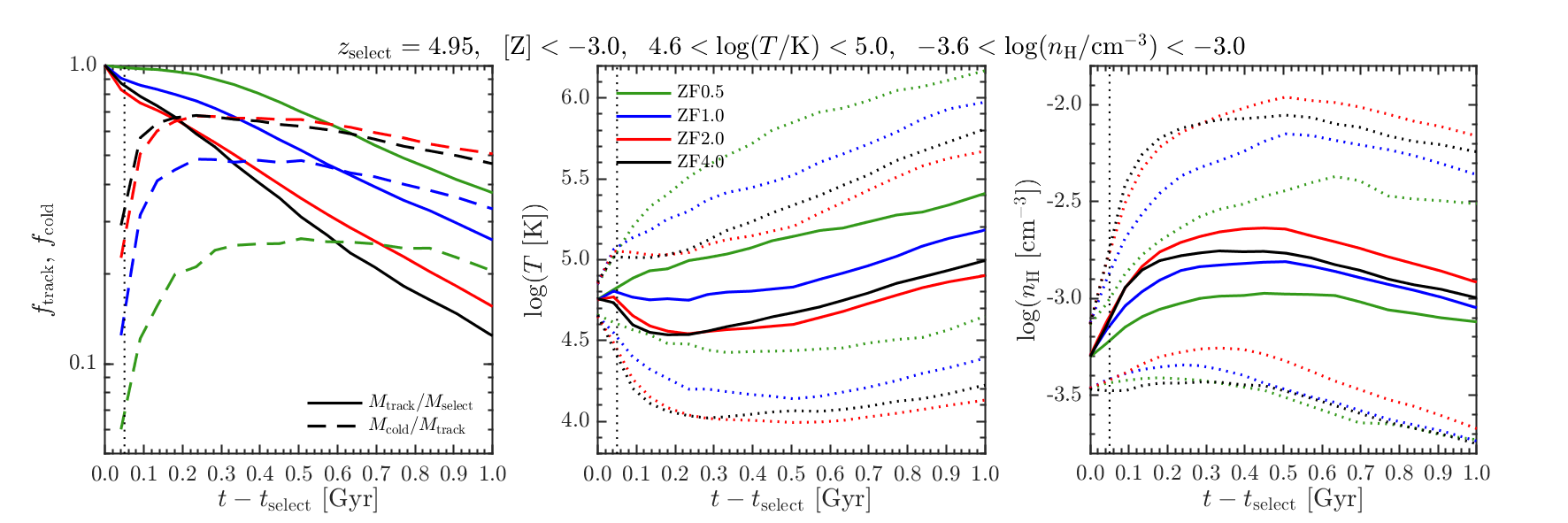}
\end{center}
\caption{Evolution of IPM gas, with $[Z]<-3.0$, selected at redshift $z\sim 5$ to have temperatures and densities in the range $4.6<{\rm log}(T/{\rm K})<5.0$ and $-3.6<{\rm log}(n_{\rm H}/\cmc)<-3.0$, in the ``island of isochoric thermal stability''. Different colours represent different simulation resolutions as marked. For each simulation, we select all IPM cells in this region of $nT$ space, and track their densities and temperatures forward in time until the cells become refined or derefined, at which point they can no longer be tracked. In each panel, the $x$ axis shows the time since the gas was selected, and the vertical dotted line marks the average cooling time of the initially selected gas, $\sim 50\Myr$. \textit{On the left}, we show the fraction of the initially selected gas mass that can still be tracked (solid lines), and the fraction of currently tracked gas mass that has cooled to $T<10^{4.5}\K$ (dashed lines). \textit{In the centre column}, we show the mean (solid) and standard deviation (dotted) of temperature (mass-weighted average of ${\rm log}(T)$). \textit{On the right}, we show the mean and standard deviation of the density (mass-weighted average of ${\rm log}(n_{\rm H})$). When the gas cells are selected, the mean and standard deviation of both $T$ and $n_{\rm H}$ are matched at all resolutions, so any differences in the evolution are not due to biases in the initial distribution of densities and temperatures. The gas cools more rapidly in higher resolution simulations, reaching lower temperatures and higher densities. As this gas cools and becomes denser, it is more likely to be refined, preventing us from tracking it further. Thus, the coldest and densest cells constantly get removed from our sample, which is why the mean temperature seems to increase even in ZF4.0, though the sharp drop in temperatures is evident from looking at one standard deviation below the mean (lower dotted lines). While ZF2.0 and ZF4.0 seem converged, with $\sim (50-70)\%$ of the tracked gas having $T<10^{4.5}\K$, ZF1.0 and ZF0.5 exhibit noticeably less cooling, with only $\sim 35\%$ and $\sim 20\%$ of the tracked gas having $T<10^{4.5}\K$, respectively.
}
\label{fig:Temp_history}
\end{figure*}

\smallskip
We now demonstrate the expected effect of our model, namely that gas cells beginning in the isochorically stable regime near the equilibrium state are prevented from cooling if $l_{\rm cool}$ is unresolved, while they do cool if $l_{\rm cool}$ is resolved. In \fig{Temp_history}, we show the thermal histories of all gas cells in the IPM with temperatures and densities in the range $T=(10^{4.6}-10^5)\K$ and $n_{\rm H}=(10^{-3.6}-10^{-3.0})\cmc$ at $z\sim 5$. Note that this is a smaller range of temperatures and densities than those highlighted in \fig{phase_low}, 
and was selected such that the mass-weighted variance of both the density and temperature in this region are the same at all resolutions. In \texttt{AREPO}, gas cells move with the flow, and approximately represent individual Lagrangian mass elements\footnote{While the mesh motion reduces mass fluxes in and out of cells, and thus for some time the tracking of a mesh generating point gives an approximation to the trajectory of the gas mass element, this is not as exact as it would be in an SPH simulation, where particles represent individual mass elements and tracking can be done unambiguously. In our case, the tracking is expected to become less accurate with time. However, averaging over many cells as we do here should preserve the mean trends, especially in the early stages, and especially given the large differences seen between different resolutions.}. Each gas cell has a unique ID and can be tracked until it is either refined or derefined, which happens when the mass of the cell is larger or smaller than the target mass by more than a factor of 2, or when the cell shape becomes too irregular. Once this happens, the cell ID is lost forever and the cell can no longer be tracked. 

\smallskip
The solid lines in the left-hand panel of \fig{Temp_history} show the mass fraction of selected cells that can still be tracked as a function of time since their selection. Clearly, this fraction decreases with time at all resolutions and redshifts. However, it decreases much faster in ZF4.0 and ZF2.0 (which behave very similarly) than in ZF1.0 or ZF0.5. After $\sim 100\Myr$, only $\sim 75\%$ of the initially selected gas mass can still be tracked in ZF4.0 and ZF2.0, compared to $\sim 100\%$ of the selected gas mass in ZF0.5. The dashed lines in this panel show the fraction of tracked mass which has $T<10^{4.5}\K$, such that it has cooled far from the quasi-equilibrium state. These fractions saturate after $\sim 100\Myr$ at $\sim 60\%$ in ZF4.0 and ZF2.0, $\sim 45\%$ in ZF1.0, and $\sim 25\%$ in ZF0.5. We note that these fractions are comparable to the mass fraction of cells where $l_{\rm cool}$ is resolved (\fig{Mwarm}). The roughly constant fraction of tracked mass at low temperatures suggests that most of the cells that can no longer be tracked are cells that have cooled and generated a cooling flow onto them from the surrounding gas, thus increasing in mass until they are refined. This is expected in runaway isobaric cooling \citep{Das21}. 

\smallskip
The center panel of \fig{Temp_history} shows the mass-weighted mean and standard-deviation of ${\rm log}(T)$ for the cells that can still be tracked as a function of time since their selection, while the right-hand panel shows the same for ${\rm log}(n_{\rm H})$. After roughly $t_{\rm cool}\sim 50\Myr$, the mean temperature in ZF4.0 and ZF2.0 begins dropping, before rising again after $\sim 300\Myr$, a result of the fact that many cells that have cooled can no longer be tracked, as described above. This is further supported by the $1-\sigma$ lower bound of the temperature, which begins dropping immediately, reaches $\sim 10^4\K$ after $100\Myr$, and remains roughly constant for $\lsim 1\Gyr$. On the other hand, the mean temperature in ZF1.0 and ZF0.5 never drops below its initial value, while the $1-\sigma$ lower bounds never fall below $10^{4.2}\K$ and $10^{4.5}\K$ respectively. Comparing the $1-\sigma$ lower bound of ${\rm log}(T)$ with the $1-\sigma$ upper bound of ${\rm log}(n_{\rm H})$, we see that for ZF4.0 and ZF2.0, the two are roughly inversely proportional to each other for the first $(100-200)\Myr$ of evolution, suggesting that the most rapidly cooling cells in these simulations maintain pressure equilibrium and do not undergo a strong compression shock during this time. 

\smallskip
We stress that the initial variance of both density and temperature among the selected cells are the same at all resolutions, as can be seen by the dotted lines at $(t-t_{\rm select})=0$ in the center and right-hand panels. This suggests that, unlike the phenomenon identified by \citet{Hummels19}, the additional cooling in ZF2.0 and ZF4.0 compared to ZF1.0 and ZF0.5 is \textit{not} simply a consequence of higher resolution simulations probing initially higher densities with shorter cooling times in a given volume. Rather, here we have selected cells with \textit{the same} distribution of initial densities, temperatures, and cooling times (see \fig{cell_low}), though not necessarily in exactly the same physical region, and found that cooling appears suppressed in low resolution simulations. This supports our hypothesis that if isochoric cooling modes are stable, but $l_{\rm cool}$ is unresolved so isobaric modes cannot grow, cooling will be artificially suppressed.


\smallskip

\smallskip
Similarly selecting cells at $z\sim 4$ yields very similar results, though the suppression of cooling in ZF1.0 and ZF0.5 is slightly less pronounced. However, at $z\sim 3$, all resolutions show much better convergence. The mean and $1-\sigma$ upper and lower bounds of both temperature and density, as well as the tracked mass fraction, are all nearly identical, though slightly more cooling is evident in ZF2.0\footnote{ZF4.0 was stopped shortly after $z\sim 3$ (\tab{sims}), so the highest resolution we can track beyond $z\sim 3$ is ZF2.0. However, since ZF2.0 and ZF4.0 appear converged at $z\sim 5$ and $z\sim 4$, we expect the same to be true at $z\sim 3$.}. 

\subsubsection{Resolving Compression due to Radiative Shock-Fronts}

\smallskip
We have argued above that the necessary physical scale to resolve is the cooling length of $\sim 10^5\K$ gas, $l_{\rm cool}=c_{\rm s}t_{\rm cool}$, where $c_{\rm s}$ and $t_{\rm cool}$ are evaluated at $T\sim 10^5\K$. However, this happens to be numerically very similar to another important scale which likely plays an important role in the thermodynamics of the IPM, namely the cooling length in the shock front, $l_{\rm cool,shock}=v_{\rm shock}t_{\rm cool,shock}$. Depending on the stage of IPM evolution, $v_{\rm shock}$ may refer either to the velocity of the original shock resulting from the sheet collision that leads to the formation of most of the HI (see \fig{MHI}), or to shocks generated by supersonic turbulence in the IPM (see \fig{profiles}). $t_{\rm cool,shock}$ refers to the cooling time in the post-shock region. If the post-shock gas has a temperature of $T_{\rm shock}\sim 10^5\K$, which is the case in the IPM in our simulations (see Appendix \se{app}), then $l_{\rm cool,shock} = \mathcal{M}_{\rm shock} l_{\rm cool}$, with $\mathcal{M}_{\rm shock}$ the shock Mach number. In our simulations, $l_{\rm cool,shock}\sim (2-3)l_{\rm cool}$.

\smallskip
While these two scales are numerically similar, their physical significance is different. $l_{\rm cool}$ represents the largest scale perturbation that can cool isobarically from a quasi-equilibrium state, while $l_{\rm cool,shock}$ represents the width of the cooling layer behind a radiatively cooling shock. In a strong radiative shock, the gas is compressed by much larger ratios than the classic adiabatic limit of 4, approaching a limiting value of the Mach number squared in an isothermal shock. However, if the cooling length behind the shock is unresolved, then the cooling time behind the shock will be artificially extended, ``locking in'' thermal energy which should have been removed \citep[e.g.][]{Yirak10}. This reduces gas compression, subsequent cooling and HI formation. If this cooling layer is well-resolved, it has been shown that thermal instabilities within it lead to the formation of small cloudlets with size of order $l_{\rm cool,shock}$ \citep{Koyama02,Heitsch06,VS06}, though these are suppressed if the cooling layer is unresolved. It has further been suggested that thermal, thin-shell and Kelvin Helmholtz instabilities in the post-shock cooling layer drive turbulence in the post-shock medium \citep{Koyama02,Heitsch06,VS06}, which may explain the reduced velocity dispersion in ZF0.5 and ZF1.0 compared to higher resolution simulations (\fig{profiles}). This creates a runaway effect, since the reduced velocity dispersion leads to less subsequent shock compression and less subsequent cooling. 

\smallskip
In summary, both $l_{\rm cool}$ in islands of isochoric stability near quasi-equilibrium thermal states, and $l_{\rm cool,shock}$ are scales necessary to resolve in order to properly model the evolution of dense, cold gas in a multiphase medium. In the IPM, these scales are roughly equal, so the two convergence criteria are the same. A detailed understanding of which mechanism is more important for the formation of cold, dense, neutral gas in the IPM is difficult to achieve using these cosmological simulations, and is left for future work using idealized simulation setups which can study both of these processes in detail.

\subsubsection{Cold Gas Survival in Turbulent Environments}

\smallskip
A last piece of physics to keep in mind is that not just the production, but the {\it survival}, of cold gas, is scale (and therefore resolution) dependent. The IPM is a highly turbulent medium, and cold gas which forms via thermal instability is rapidly broken down by turbulence. Indeed, even in the absence of ``shattering'' by thermal pressure gradients \citep[as in][]{McCourt18}, turbulence will fragment cold gas into a spectrum of small pieces and attempt to mix it with hot gas. Recently, Gronke et al. (2021, in preparation) simulated cold gas survival in a turbulent medium and found that cold gas survival requires fragments broken up by turbulence to be well resolved and remain larger than $r_{\rm crit} \sim v_{\rm turb} t_{\rm cool, mix}$, where $t_{\rm cool,mix}$ is the cooling time at $T_{\rm mix} \sim (T_{\rm c} T_{\rm h})^{1/2} \sim 10^5$K in our case, with $T_{\rm c}$ and $T_{\rm h}$ the temperatures of the cold and hot phases (see \fig{phase_low}). This is comparable to both $l_{\rm cool}$ and $l_{\rm cool,shock}$ discussed above. Gronke et al. (2021, in preparation) found considerable stochasticity and resolution dependence if clouds fragmented to scales comparable to $r_{\rm crit}$, with clouds no longer surviving in low resolution simulations, presumably because if such a cloud is resolved by a single cell, turbulence mixes it in its entirety all at once, without leaving behind a core onto which new cold gas can condense. In future work, it would be interesting to consider the resolution dependence of cold gas survival in an IPM-like environment.

\subsection{Additional Physical and Numerical Effects}
\label{sec:disc_numer}

\smallskip
In addition to the physical significance of resolving $l_{\rm cool}$ at $T\sim 10^5\K$ discussed in \se{disc_phys}, there are several general numerical considerations that affect the formation of cold, dense, neutral gas in simulations, which are not tied to a specific length scale. We discuss these below.

\subsubsection{Probing the High End of the Density PDF}

\smallskip
As discussed in Appendix \se{app}, a turbulent medium leads to a log-normal distribution of densities. Higher resolution allows us to better sample the high-density tail of this distribution. Since the cooling rate scales as the density squared, gas occupying this high-density tail will cool much faster than gas at the mean density. This was identified by \citet{Hummels19} as one of the main reasons higher resolution simulations produce much more cold gas in the CGM. This effect is certainly relevant in our simulations as well. However, it cannot be the whole story. As we showed in \fig{Temp_history}, even when we select gas with the same initial distribution of density and temperature in simulations of different resolutions, there is more cooling in higher resolution simulations. Moreover, unlike our criterion of resolving $l_{\rm cool}$ in the isochorically stable regime at $T\lsim 10^5$ as a necessary condition for forming a multiphase and shattered medium, there is no length-scale associated with this convergence criterion. Rather, convergence here requires enough cells to properly sample the density PDF up to the threshold density for SF.

\subsubsection{Numerical Mixing}

\smallskip
It is well known that Eulerian grid-codes tend to be overly diffusive on the grid-scale. This can lead to numerical mixing of hot and cold gas near interfaces between them, creating a layer of warm gas. If a cold cloud within a hot medium is poorly resolved, such that its size is only a few resolution elements, this results in artificial evaporation of the cold cloud. This process was identified by \citet{Hummels19} as one of the main reasons low resolution simulations produced less cold, neutral gas in the CGM compared to higher resolution simulations. However, while \citet{Hummels19} performed their simulations with an Eulerian AMR code, we here use a moving-mesh code. Moving-mesh codes significantly reduce artificial mixing and diffusion across the cell boundaries, because the cells move with the flow thereby reducing the fluxes travelling between them. For example, tests of the classic Kelvin-Helmholtz and Rayleigh-Taylor instabilities on a fixed- and moving-mesh with comparable resolution show significantly less artificial mixing in the moving-mesh \citep{Springel10}. Likewise, shock-tube and explosion tests with fixed- and moving-mesh codes of comparable resolution show moving-mesh codes to be far less diffusive and converge faster to the analytic solutions \citep{Springel10}. We therefore suspect that, while still present to some degree, artificial mixing on the grid scale does not play as significant a role in our results as it likely did in \citet{Hummels19}.

\subsubsection{Self-Shielding}

\smallskip
As mentioned in \se{results_1}, following \fig{nH_slice}, the shattering into small dense clumps in ZF2.0 and ZF4.0 leads to a runaway effect in HI production, due to the way self-shielding is implemented in the simulation following \citet{Rahmati13}. At $n_{\rm H}\sim 10^{-3}$, the typical density in ZF0.5 and in the quasi-equilibrium thermal state deiscussed in \se{disc_phys}, the UVB is roughly $\sim 90\%$ of its unshielded value, while at $10^{-2}\cmc$, the typical density of shattered clumps in ZF4.0, the UVB is only $10\%$ of its unshielded value. However, as we pointed out in \citet{vdv19} in our study of enhanced CGM refinement, the \citet{Rahmati13} self-shielding correction was derived from simulations with much lower resolution and in different environments, and may not be applicable to the small clouds we resolve here. To test the effect of this, \citet{vdv19} ran additional simulations without self-shielding, finding a reduction in the column densities of $\NHI\gsim 10^{16}\cms$ systems, though no qualitative change to their results regarding the lack of convergence of HI column densities with resolution. Furthermore, in our case, most of the difference in the distribution of $\NHI$ is in the optically thick regime, $\NHI>10^{17.2}\cms$ (\fig{NHI_cover}). Regardless of any sub-grid implementation, such clouds should be self-shielded, at least partially. We therefore do not expect our implementation of self-shielding to qualitatively affect our results. However, future studies employing full radiative transfer should examine the effect of self-shielding in well-resolved shattered systems like those studied here, and test the robustness of the normalization of self-shielding and HI production in such systems.

\subsection{Why are Filaments Better Converged than the IPM?}
\label{sec:disc_Fil_IPM}

\smallskip
The properties of dense, neutral gas show no sign of convergence in our simulations at any redshift, in either the IPM or filaments. However, as shown in \se{results_2}, the differences are somewhat milder in filaments than in the IPM. This is surprising, since the cooling length in filaments is smaller given their higher densities and lower temperatures.  
While this should be studied in more detail in future work, we offer here two potential explanations. First, since filaments are inherently denser than the IPM, they are more self-shielded even prior to the onset of thermal instabilities. Thus, the enhancement of self-shielding during thermal shattering and condensation has less of an effect in filaments than in the IPM, resulting in smaller differences with resolution. Second, filaments are more affected by galaxy formation feedback and winds than the IPM, since all halos with $\Mv>10^9\Msun$ lie within filaments. This can affect gas cooling in two ways. First, cold gas ejected from galaxies due to winds, or stripped from galaxies due to tidal or ram pressure stripping, can seed condensation and enhance cooling out of the ambient hot medium even if it would otherwise be thermally stable (see \citealp{Nelson20} for a discussion of this effect in the context of the CGM). Second, galactic winds and galaxy interactions within filaments generate turbulence on small scales in addition to the large-scale turbulence driven by gravitational collapse. Indeed, the typical turbulent velocities measured in filaments are larger than those in the IPM by a factor of a few. This additional turbulence compresses the gas to larger densities resulting in more efficient cooling, and also enhances the mass fraction of gas with large density fluctuations, $\delta \rho/\rho>1$, which can cool even when $l_{\rm cool}$ is unresolved in regions of isochoric stability \citep{Das21}.

\smallskip
Of course, all of the arguments presented above are even more relevant for the CGM than filaments, as gas in the CGM is both denser and closer to galaxies than gas in intergalactic filaments. The convergence properties of the CGM are thus also expected to be better than the IPM, which may be consistent with the results of several recent convergence studies of the CGM (\citealp{Peeples19,Hummels19,Suresh19}, though see \citealp{vdv19} for a larger effect). This highlights and strengthens one of our initial motivations for focusing on the IPM, namely that it offers a cleaner test of the convergence properties of thermal instabilities in a cosmological setting, allowing us to isolate and understand interesting physical effects that are otherwise diluted by uncertain galaxy formation physics.

\smallskip
Phrased more generally, the above arguments suggest that uncertain ``galaxy formation physics'' may have an advantage - their associated energy input can regulate, to some extent, thermal fragmentation and shattering, and in this sense acts as a form of numerical closure. On the other hand, the more or less pristine IPM represents a pure ab-initio experiment, which is not regulated by additional small-scale physics and their associated energy input. Without such regulation, it is much more difficult to obtain a quantitatively converged numerical answer for a problem with a dynamic range as large as the IPM. Put another way, the simulated physics in the IPM are relatively simple but the numerics are hard, while for the dense filaments and the CGM regulation by star formation may stabilize the numerics to some extent, but the physics are more uncertain. While we cannot say for certain at this time that filaments are regulated by feedback from galaxies within them, this is an intriguing possibility worth exploring further.

\section{Summary and Conclusions} 
\label{sec:conc}

\smallskip
Using a novel suite of simulations zooming in on an intergalactic sheet or ``pancake'' between two massive halos, we performed an in-depth convergence study of the thermal properties and HI content of the WHIM at redshifts $z\sim 3-5$. During this epoch, a strong accretion shock forms around the main pancake following a collision between two smaller sheets at $z\sim 5$. Gas in the post-shock region proceeds to rapidly cool, leading to thermal instabilities and the formation of a multiphase medium. Our lowest resolution simulation has a gas cell mass of $\sim 7.7\times 10^6\Msun$, comparable to Illustris TNG300, while our highest resolution simulation has a gas cell mass of $\sim 1.5\times 10^4\Msun$, $\sim 8$ times better than Illustris TNG50. To focus on the IGM, we removed all gas associated with halos containing at least 32 dark matter particles. We then separated intergalactic filaments, within which all halos with $\Mv>10^9\Msun$ reside, from the intra-pancake medium, or IPM, in between the filaments and far from any star-forming galaxies. This separation was performed based on metallicity, with filaments having $Z>10^{-3}Z_{\odot}$ and the IPM having lower metallicity values. In addition to studying the convergence of HI mass, morphology, and distribution, and the gas thermal properties in general, we identified several physical and numerical effects governing the convergence. Our main results can be summarized as follows:

\begin{enumerate}

    \item Increasing the resolution results in noticeably more HI in both filaments and the IPM. Large-scale maps reveal an increase in both the overall normalization and degree of fluctuations in $\NHI$ (\fig{NHI_large}), which is also reflected in the total HI mass within the filaments and the IPM (\fig{MHI}) and the covering fractions of $\NHI>10^{15}\cms$ absorbers (\fig{NHI_cover}). This reflects the fact that the IGM, and in particular the IPM, has a shattered structure in high resolution simulations, consisting of small $\sim \kpc$ scale dense clouds which are absent in lower resolution simulations (\fig{nH_slice}). Most of the HI in the IPM is concentrated in these clouds (\fig{nH_slice}), which have very high neutral fractions (\fig{xHI_hist}). As the resolution is increased, these clouds become more prevalent, resulting in large clumping factors that are far from converged in both filaments and the IPM (\fig{clumping}). 
    
    \item In our highest resolution simulation, the covering fraction of LLSs in the IPM with $Z<10^{-3}Z_{\odot}$ increases from $\sim 3\%$ at $z\sim 3$ to $\sim 30\%$ at $z\sim 5$ (\fig{NHI_cover}, \tab{covering}). During the same period, the covering fraction of sightlines with clumping factor greater than 10 increases from $\sim 10\%$ to $\sim 30\%$ (\fig{clumping}, \tab{covering}). We also detect DLAs with $\NHI>10^{20}\cmc$. These are most commonly found in intergalactic filaments, where their covering fraction increases from $\sim 0.6\%$ at $z\sim 3$ to $\sim 5\%$ at $z\sim 5$. At $z\sim (4-5)$, the IPM contains DLAs as well, with a covering fraction of order $\sim 1\%$ (\tab{covering}). These intergalactic DLAs may represent the population of ``missing'' low-metallicity DLAs, with $[Z]<-2.0$, highlighted by \citet{Stern21} 
    
    \item While none of the aforementioned properties are converged in either filaments or the IPM, convergence is slightly better in the filaments. This is in part due to the fact that at $z>3$ the filaments are dense enough to be self-shielded from the UV background even in low resolution simulations, and in part due to the fact that they are more susceptible to small scale perturbations from the galaxies that lie within them, which can seed additional cooling even in an otherwise stable system.
    
    \item While the distributions of density and temperature are also unconverged, the pressure distribution is well converged in both filaments and the IPM (\fig{dens_low}). 
    
    \item Focusing on the IPM, as the resolution is lowered the mass of gas with $T\sim (10^{4.6}-10^5)\K$ significantly and systematically increases, while the mass of gas with $T< 10^{4.5}\K$ decreases (\figs{phase_low}-\figss{dens_low}). This creates the effect of a cooling ``bottleneck'' where the efficiency of cooling in $T\lsim 10^5\K$ gas is reduced in low resolution simulations. This bottleneck is correlated with the fraction of mass in this temperature range where the cooling length, $l_{\rm cool}=c_{\rm s}t_{\rm cool}$ is resolved (\figs{cell_low}-\figss{Mwarm}). At $z\sim 3$, this bottleneck seems to have ``opened up'' as $l_{\rm cool}$ is resolved in most of the gas mass. 
    
    \item By tracing the thermal histories of individual gas cells in the region of $nT$ space where the ``pile-up'' in low resolution simulations was most evident, we find that cells in low resolution simulations do not cool as much or as fast as those in high resolution simulations (\fig{Temp_history}). This is despite the fact that the cells were selected to have the same initial distribution of densities, temperatures, and cooling times in all resolutions, and is suggestive that some physical process prevents the low resolution gas from cooling. The excess of cold gas mass in high resolution simulations thus does \textit{not} appear to be simply a result of higher resolution simulations probing higher densities with shorter cooling times. Rather, we propose that this region of $nT$ space is stable to isochoric thermal instability, while isobaric thermal instability can only grow on scales $\lsim l_{\rm cool}$. If this scale is unresolved, thermal instabilities and cooling become highly inefficient and must rely on non-linear perturbations. Resolving this scale is also important as it is roughly the expected width of the cooling layer behind shocks surrounding and within the sheet. 
    
    \item While resolving $l_{\rm cool}$ at $T\lsim 10^5\K$ appears a necessary condition for the formation of a multiphase medium in the IPM, or indeed any pressure-confined shock-heated medium, it is not sufficient for convergence of the HI morphology and clumpiness at the highest column densities. These remain unconverged even in our highest resolution simulation. We suspect that in order for these results to converge in the absence of additional physics such as thermal conduction or diffusion, simulations must resolve $l_{\rm cool,min}$, the minimal cooling length at $T\gsim 10^4\K$ \citep{McCourt18}. This is sub-kpc in the IPM and tens of pc in the CGM, beyond the current capabilities of cosmological simulations. 
    Future work using idealized simulations should study the convergence properties of multiphase media at the scale of $l_{\rm cool,min}$ and below.
    
\end{enumerate}

\smallskip
The results presented here show that the IGM, and in particular the IPM, has a multiphase structure resulting from thermal instabilities which are unresolved in standard cosmological simulations. This is very similar to the situation in the CGM, as found by several recent studies. Studies attempting to compare cosmological simulations to observations of strong HI absorption systems must be aware of this, and that such systems may be unresolved, especially at low metallicities of $[Z]\lsim -2.0$. Our study also highlights the usefulness of the IPM for studying thermal instabilities and the formation of a multiphase medium through shattering in a cosmological context, without the complicating factors of uncertain galaxy formation physics, since the regions we have focused on are so far away from any star-forming galaxies. This has allowed us to gain greater insight into the causes of the lack of convergence. Future work should focus on two main avenues. First, the processes identified here as potential causes for the lack of convergence and its relation to the cooling length of $\lsim 10^5\K$ gas should be explored in detail using idealized numerical experiments. Second, the cosmological frequency of such systems should be estimated in order to ascertain the significance of these processes for cosmological surveys.



\section*{Acknowledgments} 
We thank Prakriti Pal Choudhury, Hitesh Kishore Das, Frederick Davies, Max Gronke, Joseph F. Hennawi, Cameron Hummels, and Prateek Sharma for very helpful discussions, and comments on earlier drafts of this work. NM also thanks Friedrich Roepke for supporting continued use of the computing facilities at HITS past the expiration of his fellowship there. This work was partly funded by the Klauss Tschira Foundation through the HITS Yale Program in Astropysics (HYPA). NM acknowledges support from the Gordon and Betty Moore Foundation through Grant GBMF7392 and from the National Science Foundation under Grant No. NSF PHY-1748958. FvdB is supported by the National Aeronautics and Space Administration through Grant No. 19-ATP19-0059 issued as part of the Astrophysics Theory Program. FvdV is supported by a Royal Society University Research Fellowship.

\begin{appendix}

\section{A. Isochorically Stable Thermal Equilibrium in the IPM}
\label{sec:app}


\begin{figure*}
\begin{center}
\includegraphics[trim={0.1cm 0.0cm 0.0cm 0.0cm}, clip, width =0.98 \textwidth]{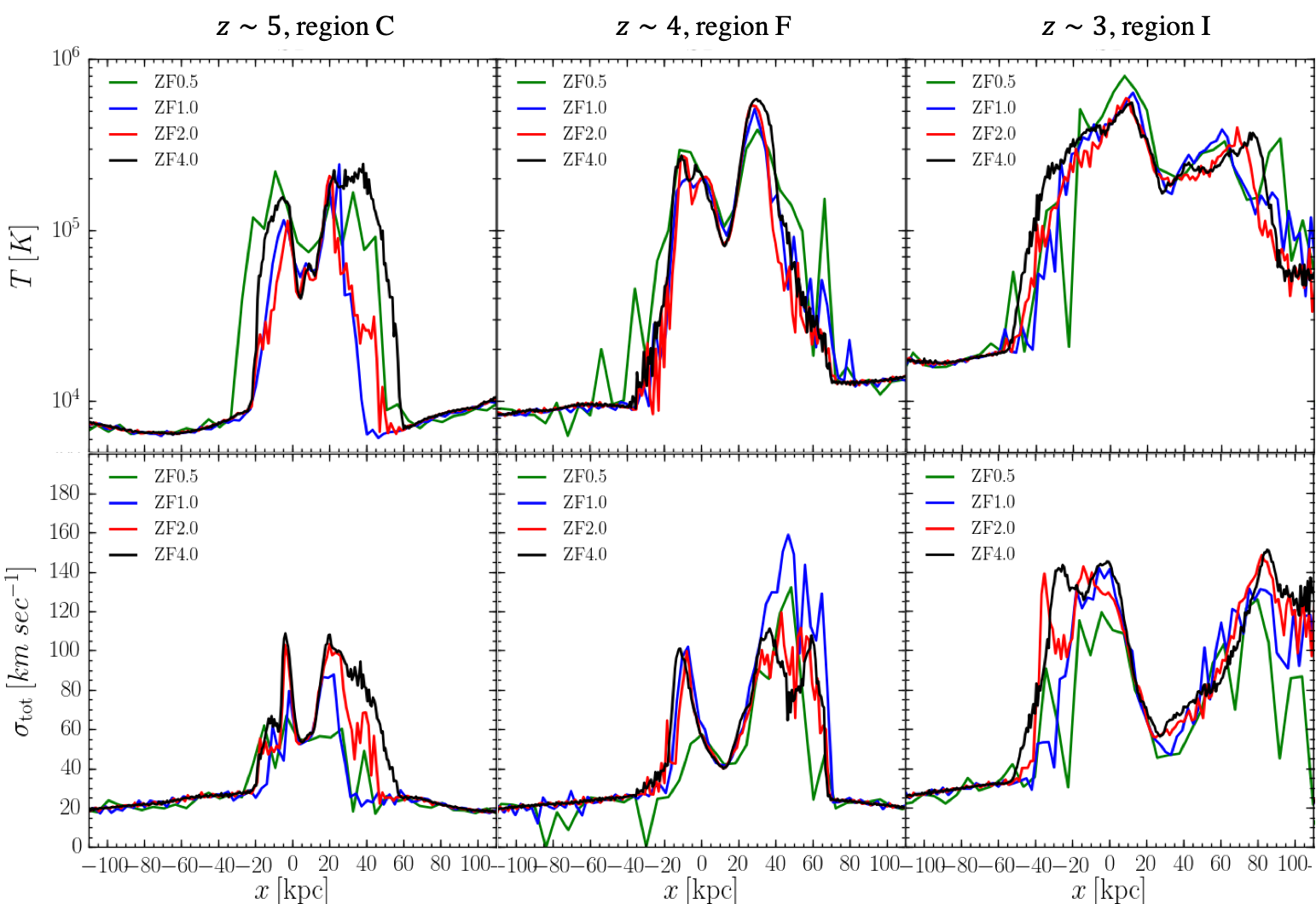}
\end{center}
\caption{Profiles of gas temperature (top) and 3d velocity dispersion (bottom) along the axis perpendicular to the sheet plane. We show profiles in the three regions highlighted in \fig{NH_large}, region C at $z\sim 5$ (left), region F at $z\sim 4$ (center, also shown in \fig{nH_slice}), and region I at $z\sim 3$. Different colors correspond to different resolutions, as marked. The bin size used along $x$ when evaluating the profiles is representative of the typical cell sizes in the different simulations, $0.75$, $1.5$, $3.0$, and $6.0\kpc$ for ZF4.0, 2.0, 1.0, and 0.5 respectively. The temperature profiles are mass-weighted, and the velocity dispersion is the sum in quadrature of the three components of the mass-weighted velocity dispersion (see text). The shocks above and below the sheet are clearly visible in both the temperature and velocity dispersion profiles. The post shock temperature rises from $\sim 10^5\K$ at $z\sim 5$ to $\sim 6\times 10^5\K$ at $z\sim 3$, while the midplane temperature rises from $\sim 0.6\times 10^5\K$ to $\sim 2\times 10^5\K$ in the same timespan. The velocity dispersion in the post-shock region rises from $\sim 100\kms$ to $\sim 140\kms$ from $z\sim 5$ to $z\sim 3$, while the midplane value is $\sim (50-60)\kms$ throughout. The eddy turnover times, evaluated from the peak to the minimum of the velocity dispersion profiles, are very similar to the typical cooling times of IPM gas in the region of $nT$ space highlighted in \fig{phase_low}, 
while the dissipated energy can maintain temperatures of order $\sim 10^5\K$. This suggests that gravity-driven turbulent dissipation within the post-shock sheet can maintain a quasi-equilibrium thermal state in the IPM.
}
\label{fig:profiles}
\end{figure*}

\smallskip
In this section, we demonstrate the two necessary conditions for our hypothesis presented in \se{iso_stab}, namely that the ``cooling bottleneck'' in low resolution simulations at $T\lsim 10^5\K$ (see \figs{phase_low}-\figss{dens_low}), and its relation to resolving the cooling length at these temperatures (see \figs{cell_low}-\figss{Mwarm}), are due to this region being stable to isochoric perturbations while isobaric perturbations are unresolved. As stated in \se{iso_stab}, the two necessary conditions for this are (1) there exists an approximate thermal equlibrium near $T\sim 10^5\K$ in the IPM, and (2) this is in a regime where isochoric modes are stable while isobaric modes are unstable.

\smallskip
In \fig{profiles}, we show vertical profiles of temperature (top) and velocity dispersion (bottom), computed in uniform bins of $x$, which runs perpendicular to the sheet plane, in regions C, F, and I from \fig{NH_large} at redshifts $z\sim 5$, 4, and 3 respectively. We have also examined the other IPM regions marked with rectangles in \fig{NH_large} and find the results to be very similar. Recall that region F is also highlighted in \fig{nH_slice}. The bin sizes are chosen to reflect the different spatial resolutions of the different simulations, $(0.75,\,1.5,\,3.0,\,6.0)\kpc$ for ZF(4.0, 2.0, 1.0, 0.5) respectively. The temperature profiles represent the mass-weighted average temperature in the highlighted regions of the $yz$ plane, in each bin of $x$, namely 
\be 
\label{eq:Tavg}
T(x) = \left<T(x,y,z)\right>,
\ee 
where $\left<.\right>$ represents a mass-weighted average within the highlighted region of the $yz$ plane per bin of $x$. To compute the velocity dispersion, we first evaluated the three components of the mass-weighted average velocity (i.e. the center of mass velocity) in each bin of $x$ within each region. We then computed the three components of the velocity dispersion separately, and sum them in quadrature to obtain the total velocity dispersion, namely
\be 
\label{eq:sig_avg}
\sigma^2_i(x) = \left<v_i^2(x,y,z)\right>-\left<v_i(x,y,z)\right>^2,
\ee 
\be 
\label{eq:sig_tot}
\sigma^2_{\rm tot}(x) = \sum_{i=x,y,z} \sigma^2_i(x).  
\ee 

\smallskip
The shocks on either side of the sheet are visible in the temperature profiles, and it is apparent that they are not perfectly symmetric. The post-shock temperatures are $\sim (1-2)$, $(2-5)$, and $(3-6)\times 10^5\K$ at $z\sim 5$, $4$, and $3$ respectively, and are very similar at all resolutions. The average midplane temperature is $\sim 0.6$, $1.0$, and $2.0\times 10^5\K$ at $z\sim 5$, $4$, and $3$ respectively, though it is slightly higher in lower resolution simulations, particularly evident in ZF0.5 at $z\sim 5$. Overall, the temperature profiles are very similar at all resolutions, especially at $z\sim 3$ and $4$, though at $z\sim 5$ ZF4.0 shows a slightly larger shock region. 
The velocity dispersion profiles show more variance between different resolutions. ZF4.0 and ZF2.0 are largely very similar, though ZF1.0 and ZF0.5 tend to have lower values in the post-shock regions. Focusing on the ZF4.0 profiles, the shocks are clearly visible as sharp-peaks in the velocity dispersion, reaching values of $\sim 100\kms$ at $z\sim 5$ and $4$, and $\sim 140\kms$ at $z\sim 3$. The velocity dispersion rapidly declines towards the sheet midplane, reaching $\sim (50-60)\kms$ at all redshifts. It is worth noting that the post-shock values of the velocity dispersion are super-sonic with respect to the post-shock temperatures. This is made possible by the inclined nature of the sheet collision at $z\sim 5$ which generates an oblique shock. 

\smallskip
We evaluate the turbulent kinetic energy dissipated from the post-shock peak to the midplane by computing 
\be
\label{eq:disp_eff}
\sigma^2_{\rm dis} \simeq \sigma^2_{\rm peak}-\sigma^2_{\rm mid}.
\ee
{\no}We obtain $\sigma_{\rm dis}\sim 80\kms$ at $z\sim 5$ and $4$, and $\sigma_{\rm dis}\sim 125\kms$ at $z\sim 3$, corresponding to effective temperatures of $\sim 2\times 10^5\K$ and $\sim 6\times 10^5\K$ respectively. We define a characteristic eddy scale as 
\be
\label{eq:leddy}
l_{\rm eddy} \simeq 0.5(x_{\rm peak}-x_{\rm mid}),
\ee
{\no}namely half the distance between the maximum and minimum of the velocity dispersion in the sheet. We obtain $l_{\rm eddy}\sim 4$, $10$, and $25\kpc$ at $z\sim 5$, $4$, and $3$ respectively. We now define a characteristic heating time as the eddy turnover time, 
\be
\label{eq:teddy}
t_{\rm eddy} \simeq l_{\rm eddy} / \sigma_{\rm dis},
\ee
{\no}and obtain $t_{\rm eddy}\sim 50$, $100$, and $200\Myr$ at $z\sim 5$, $4$, and $3$ respectively. These are extremely similar to the typical cooling times of IPM gas in the region of $nT$ space highlighted in \fig{phase_low} 
at the respective redshifts.

\smallskip
The above analysis demonstrates the first of the two conditions mentioned above, namely that a thermal quasi-equilibrium state exists in the IPM at temperatures $T\lsim 10^5\K$, where radiative cooling is balanced by energy dissipation of supersonic turbulent motions in the post-shock medium. These turbulent motions are themselves powered by the gravitational potential energy of the large-scale sheet, which sets the velocities of the infalling gas that result in the shock. This is thus a form of ``gravitational heating'', as discussed in other contexts of how to maintain a hot CGM in massive galaxies without AGN feedback \citep[e.g.][]{Mo05,Dekel08,Birnboim11}. While this form of turbulent heating does not imply a true equilibrium where heating balances cooling for any given parcel of gas, it does imply that equilibrium can be maintained in a stochastic, ensemble-averaged sense, which we assume is enough to apply the insights of \citet{Das21}. 

\begin{figure}
\includegraphics[trim={0.3cm 0.1cm 0.2cm 0.2cm}, clip, width =0.47 \textwidth]{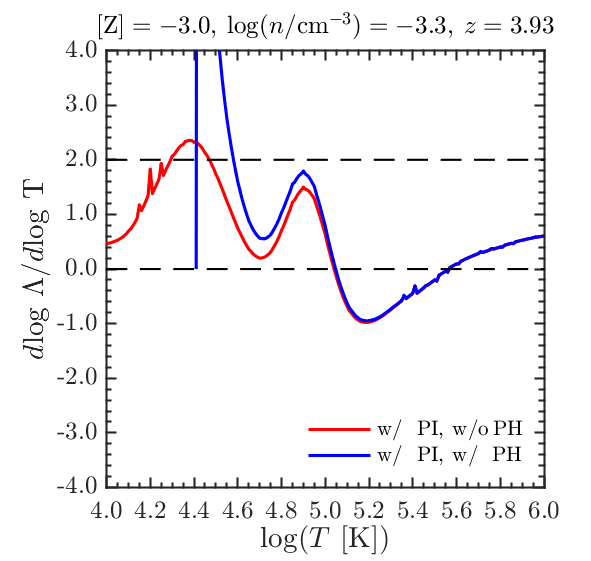}
\caption{Logarithmic derivative of the cooling function with respect to temperature, $\Lambda_{\rm T}=d{\rm log}~\Lambda/d{\rm log}~T$, as a function of temperature. We assume a metallicity of $[Z]=-3.0$, and include photoionization by the UVB, either without (red) or with (blue) accounting for photoheating as well. We assume the $z=3.93$ value of the UVB, and a gas density of $n=10^{-3.3}\cmc$. In both cases, at $T\sim (10^{4.6}-10^5)\K$, we have $0<\Lambda_{\rm T}<2$, implying that isochoric cooling modes are stable while isobaric modes are unstable.}
\label{fig:dlnL}
\end{figure}

\smallskip
We now address the second issue, namely that the quasi-equilibrium state identified above is in a regime where isochoric modes are stable while isobaric modes are unstable. The turbulent heating rate we described above has no explicit dependence on gas density or temperature, but only on position. Photoheating by the UV background has a negligible effect on gas with $T\lsim 10^5\K$. Photoheating by AGN can affect gas at these temperatures \citep{Vogelsberger13}, however the IPM is far away from any massive galaxy or accreting black hole, rendering AGN heating negligible. We therefore assume a total heating rate which is independent of density or temperature, which was the case studied by \citet{Das21}. In this case, the stability of isobaric and isochoric modes depends on the logarithmic derivative of the cooling function versus temperature, $\Lambda_{\rm T}=d{\rm log}~\Lambda/d{\rm log}~T$. If $\Lambda_{\rm T}<0$ then both isobaric and isochoric modes are unstable, if $0<\Lambda_{\rm T}<2$ then isobaric modes are unstable while isochoric modes are stable, and if $\Lambda_{\rm T}>2$ then both modes are stable \citep{Das21}. In \fig{dlnL}, we show $\Lambda_{\rm T}$ as a function of temperature for the cooling curve used in our simulations. We assume a metallicity of $[Z]=-3.0$, though the results are effectively identical for metallicity as high as $[Z]\gsim -2.0$ and for primordial gas. 
While photoheating by the UVB is expected to be negligible at $T\sim 10^5\K$, photoionization is still important and can alter the cooling curve. 
We show results assuming photoionization due to the UVB at $z=3.93$, both with and without accounting for photoheating as well in blue and red respectively. 
We here assumed a density of $n=10^{-3.3}\cmc$, the median density in the region of $nT$ space highlighted in \fig{phase_low}. 
While photoionization alters the cooling curve at all $T<2\times 10^5\K$, photoheating has hardly any effect at $T\gsim 6\times 10^4\K$ as expected. Regardless, in both cases gas with $T\sim (10^{4.6}-10^5)\K$ has $0<\Lambda_{\rm T}<2$ so isochoric cooling modes are stable. 
These results are unchanged by assuming the UVB at $z\sim 3$ or $z\sim 5$ instead of $z\sim 4$.

\smallskip
We note that the above analysis suggests that \textit{linear} isochoric perturbations around the quasi-equilibrium thermal state should be stable. However, perturbations with initially non-linear amplitudes can cool isochorically even if linear isochoric modes are stable \citep{Das21}. Supersonic turbulence tends to generate a lognormal density PDF \citep[e.g.][]{VS94,Padoan97,Scalo98,Federrath08,Price11,Hopkins12b}, as indeed seen in the IPM (\fig{dens_low}). If we assume that the density at the peak of the lognormal distribution, ${\bar{\rho}}$, corresponds to the equilibrium density, then clouds with $\delta \rho/{\bar{\rho}}<1$ will be unable to cool isochorically, while those with $\delta \rho/{\bar{\rho}}>1$ will be unaffected by the arguments presented above. For isothermal turbulence, the width of the lognormal distribution is given by 
\be 
\label{eq:turb_mach_pdf}
\sigma_{\rm ln(\rho)} \simeq \left[{\rm ln}\left(1+b_{\rm turb}^2\mathcal{M}_{\rm turb}^2\right)\right]^{1/2},
\ee
{\no}where $\mathcal{M}_{\rm turb}$ is the turbulent Mach number and $b_{\rm turb}$ depends on the ratio of compressive to solenoidal forcing driving the turbulence. While our system is not isothermal, and \equ{turb_mach_pdf} may not be precisely valid in a gravitationally stratified medium in any case \citep{Mohapatra21}, we use this as a proxy for the expected width of the PDF. To obtain an upper limit, we assume purely compressive forcing with $b=1$ \citep{Federrath08}. For a turbulent Mach number of $\mathcal{M}_{\rm turb}\sim (1-2)$ (\fig{profiles}) we obtain $\sigma_{\rm ln(\rho)} \sim (0.8-1.25)$. The condition $\delta \rho/{\bar{\rho}}<1$ corresponds to ${\rm ln}(\rho/{\bar{\rho}})<0.7$, so $\sim (40-65)\%$ of the gas mass should have $\delta \rho/{\bar{\rho}}<1$ and be affected by the isochoric stability criterion discussed above.

\end{appendix}

\bibliographystyle{mn2e} 
\bibliography{biblio}

\label{lastpage} 
 
\end{document}